\documentclass[onecolumn, a4paper, accepted=2024-02-26]{quantumarticle}

\pdfoutput=1

\usepackage{amsmath, amsfonts, amssymb, amsthm}
\usepackage{hyperref}
\usepackage{bbm}
\usepackage{physics}
\usepackage{tabularx}
\usepackage{tikz}
\usepackage[numbers, compress]{natbib}

\makeatletter
\DeclareFontFamily{OMX}{MnSymbolE}{}
\DeclareSymbolFont{MnLargeSymbols}{OMX}{MnSymbolE}{m}{n}
\SetSymbolFont{MnLargeSymbols}{bold}{OMX}{MnSymbolE}{b}{n}
\DeclareFontShape{OMX}{MnSymbolE}{m}{n}{
	<-6>	MnSymbolE5
	<6-7>	MnSymbolE6
	<7-8>	MnSymbolE7
	<8-9>	MnSymbolE8
	<9-10>	MnSymbolE9
	<10-12>	MnSymbolE10
	<12->	MnSymbolE12
}{}
\DeclareFontShape{OMX}{MnSymbolE}{b}{n}{
	<-6>	MnSymbolE-Bold5
	<6-7>	MnSymbolE-Bold6
	<7-8>	MnSymbolE-Bold7
	<8-9>	MnSymbolE-Bold8
	<9-10>	MnSymbolE-Bold9
	<10-12>	MnSymbolE-Bold10
	<12->	MnSymbolE-Bold12
}{}

\let\llangle\@undefined
\let\rrangle\@undefined
\DeclareMathDelimiter{\llangle}{\mathopen}{MnLargeSymbols}{'164}{MnLargeSymbols}{'164}
\DeclareMathDelimiter{\rrangle}{\mathclose}{MnLargeSymbols}{'171}{MnLargeSymbols}{'171}
\makeatother

\newcommand{\Bra}[1]{\left\llangle #1 \right|}
\newcommand{\Ket}[1]{\left| #1 \right\rrangle}
\newcommand{\Braket}[2]{\left\llangle #1 \vphantom{#2} \right|\kern-0.6ex\left. #2 \vphantom{#1}\right\rrangle}
\begin{document}
	\title{Emergent parallel transport and curvature in

	Hermitian and non-Hermitian quantum mechanics}

	\author{Chia-Yi Ju}
	\affiliation{Department of Physics, National Sun Yat-sen University, Kaohsiung 80424, Taiwan}
	\affiliation{Center for Theoretical and Computational Physics, National Sun Yat-sen University, Kaohsiung 80424, Taiwan}
	\orcid{0000-0001-7038-3375}
	\author{Adam Miranowicz}
	\affiliation{Institute of Spintronics and Quantum Information, Faculty of Physics, Adam Mickiewicz University, 61-614 Pozna\'n, Poland}
	\affiliation{Theoretical Quantum Physics Laboratory, Cluster for Pioneering Research, RIKEN, Wakoshi, Saitama, 351-0198, Japan}
	\orcid{0000-0002-8222-9268}
	\author{Yueh-Nan Chen}
	\affiliation{Department of Physics, National Cheng Kung University, Tainan 70101, Taiwan}
	\affiliation{Center for Quantum Frontiers of Research \& Technology, NCKU, Tainan 70101, Taiwan}
	\affiliation{Physics Division, National Center for Theoretical Sciences, Taipei 10617, Taiwan}
	\orcid{0000-0002-2785-7675}
	\author{Guang-Yin Chen$^{\dagger}$}
	\email{gychen@phys.nchu.edu.tw}
	\affiliation{Department of Physics, National Chung Hsing University, Taichung 40227, Taiwan}
	\orcid{0000-0002-0075-6428}
	\author{Franco Nori}
	\affiliation{Theoretical Quantum Physics Laboratory, Cluster for Pioneering Research, RIKEN, Wakoshi, Saitama, 351-0198, Japan}
	\affiliation{Quantum Computing Center, RIKEN, Wakoshi, Saitama, 351-0198, Japan}
	\affiliation{Physics Department, University of Michigan, Ann Arbor, MI 48109-1040, USA}
	\orcid{0000-0003-3682-7432}

	\begin{abstract}
		Studies have shown that the Hilbert spaces of non-Hermitian systems require nontrivial metrics. Here, we demonstrate how evolution dimensions, in addition to time, can emerge naturally from a geometric formalism. Specifically, in this formalism, Hamiltonians can be interpreted as a Christoffel symbol-like operators, and the Schr\"{o}dinger equation as a parallel transport in this formalism. We then derive the evolution equations for the states and metrics along the emergent dimensions and find that the curvature of the Hilbert space bundle for any given closed system is locally flat. Finally, we show that the fidelity susceptibilities and the Berry curvatures of states are related to these emergent parallel transports.
	\end{abstract}

	\maketitle

	\section{Introduction}

		Since the discoveries and development of $\cal{PT}$-symmetric~\cite{Bender1998, Bender2007, Makris2008, ElGanainy2018} and pseudo-Hermitian~\cite{Mostafazadeh2003, Mostafazadeh2010} quantum mechanics (QM), non-Hermitian QM has become one of the major research fields~\cite{Peng2014, Jing2014, Bender2016, Bender2017, Miller2017, Leykam2017, Quijandria2018, ElGanainy2019, Liu2019, Ge2019, Parto2020, Ashida2020, Cirio2022, Bergholtz2021, Zhang2022, Fring2022, Fang2022, Chen2022, Fring2023, Znojil2024}. Most of the non-Hermitian studies focus on generalizing Hermitian QM to non-Hermitian QM~\cite{Znojil2008, Znojil2009, Brody2013} or finding some exotic properties of non-Hermitian quantum systems~\cite{Hodaei2017, Bliokh2019, Znojil2020, Znojil2021}. However this study is performed in the opposite direction. Specifically, we extend the geometric (i.e., fiber-bundle) formalism inspired by non-Hermitian QM and show that it can also be applied to Hermitian quantum systems.

		\begin{table}[h]
			\renewcommand*{\arraystretch}{1.6}
			\begin{center}
				\begin{tabular}{| c | c |}
					\hline
					Conventional QM & Non-Hermitian (Metricized) QM\\
					\hline
					$\braket{\phi}{\psi} = \begin{pmatrix}
						\phi_{1}^* & \phi_2^*
					\end{pmatrix}\begin{pmatrix}
						\psi_1\\ \psi_2
					\end{pmatrix}$ & \begin{tabular}{l}
						$\Braket{\phi}{\psi} = \bra{\phi} G \ket{\psi}$\\
						$= \begin{pmatrix}
							\phi_{1}^* & \phi_2^*
						\end{pmatrix}\begin{pmatrix}
							g_{11} & g_{12}\\
							g^*_{12} & g_{22}
						\end{pmatrix}\begin{pmatrix}
							\psi_1\\ \psi_2
						\end{pmatrix}$
					\end{tabular}\\
					\hline
				\end{tabular}
			\end{center}
			\caption{Comparison of two kinds of inner products in a two-dimensional Hilbert space. The dual state in the conventional QM is just the Hermitian conjugate of the state; the dual state in the metricized QM carries an additional structure, namely, the metric operator $G$. Note that in Hermitian QM, the $G$ can always be chosen to be the identity, which reduces back to conventional QM.}
			\label{ComapringInnerProducts}
		\end{table}

		Strong evidence is suggesting that the Hilbert space bundles, where the fiber is a Hilbert space and the base space is time, of closed non-Hermitian quantum systems have some nontrivial geometric structures~\cite{Mostafazadeh2003, Bender2004, Mostafazadeh2004, Ju2021} (see Table~\ref{ComapringInnerProducts}). It was pointed out~\cite{Ju2019} that treating Schr\"{o}dinger's equation as a parallel transport, an analogue of a less strict geodesic with the Hamiltonian being a ``generalized'' Christoffel symbol~\cite{Misner2017, Wald, DonaldStoker2019} in a fiber bundle [see Table~\ref{CompareGRandQM} and Appendix~\ref{Appendix:MQMandGR} for the analogy with general relativity (GR)], along the time evolution dimension leads~\cite{Ju2019} to a self-consistent QM, which can apply to both Hermitian and non-Hermitian quantum systems.

		\begin{table*}
			\renewcommand*{\arraystretch}{1.6}
			\begin{center}
				\begin{tabular}{| >{\centering\arraybackslash}m{0.16\textwidth} | >{\centering\arraybackslash}m{0.35\textwidth} | >{\centering\arraybackslash}m{0.37\textwidth} |}
					\hline
					& General relativity (GR) & (Non-)Hermitian (metricized) QM\hspace{-0.2cm}\\
					\hline
					Inner product & $g(U, V) = U^\mu g_{\mu \nu} V^\nu$ & $\Braket{\psi_1}{\psi_2} = \bra{\psi_1} G \ket{\psi_2}$\\
					\hline
					Field equation for the metric\vspace{0.08cm}
					& \begin{tabular}{l}
						$0 = \nabla_\lambda g_{\mu \nu}$\\
						$\quad\! = \partial_\lambda g_{\mu \nu} - \Gamma^\rho_{~\lambda \mu} g_{\rho \nu} - g_{\mu \rho} \Gamma^\rho_{~\lambda \nu}$
					\end{tabular}\vspace{0.1cm} & \raisebox{-0.05cm}{$\bigg\lbrace$}\begin{tabular}{l}
						$0 = \nabla_t G = \partial_t G - i G H + i H^\dagger G$\\
						$0 = \nabla_i G = \partial_i G - i G K_i + i K_i^\dagger G$
					\end{tabular}\vspace{0.1cm}\\
					\hline
					Field equation for vectors\vspace{0.05cm}
					& \begin{tabular}{l}
						$0 = \dfrac{d x^\nu}{d \tau} \left( \nabla_\nu \dfrac{d x^\mu}{d \tau} \right)$\\
						$\quad\! = \dfrac{d^2 x^\mu}{d \tau^2} + \Gamma^\mu_{~ \nu \lambda} \dfrac{d x^\nu}{d \tau} \dfrac{d x^\lambda}{d \tau}$
					\end{tabular}\vphantom{\rule{0pt}{1.1cm}}\vspace{0.2cm} & \raisebox{-0.05cm}{$\bigg\lbrace$}\begin{tabular}{l}
						$0 = \nabla_t \ket{\psi} = \left( \partial_t + i H \right) \ket{\psi}$\\
						$0 = \nabla_i \ket{\psi} = \left( \partial_i + i K_i \right) \ket{\psi}$
					\end{tabular}\vspace{0.1cm}\\
					\hline
					Curvature\vspace{0.08cm} & $T_{\mu \nu} = G_{\mu \nu} = R_{\mu \nu} - \dfrac{1}{2} g_{\mu \nu} R$\vphantom{\rule{0pt}{0.6cm}}\vspace{0.1cm} & $F_{ij} = 0 = F_{ti}$\vspace{0.08cm}\\
					\hline
				\end{tabular}
			\end{center}
			\caption{Comparison of the basic concepts in GR and non-Hermitian quantum mechanics. Although the inner products in both cases are affected by the geometry of the space, the vectors in GR~\cite{Misner2017, Wald, DonaldStoker2019} live in the same space of the coordinates, namely, the spacetime; but the vectors in QM are defined in a Hilbert space where the coordinates form another space. Hence, QM can only be described by a fiber bundle which corresponds to, roughly speaking, a generalized Riemannian geometry. While the equation of motion in GR follows the geodesic equation, i.e., a parallel transport along itself, the equation of motion in metricized QM is just the parallel transport along the evolving direction. The curvature in GR is determined by the external source (energy momentum), but the local curvature of the Hilbert space bundle of any closed quantum system is always zero.}\label{CompareGRandQM}
		\end{table*}

\begin{figure*}
			\begin{center}
				\includegraphics[scale=0.8]{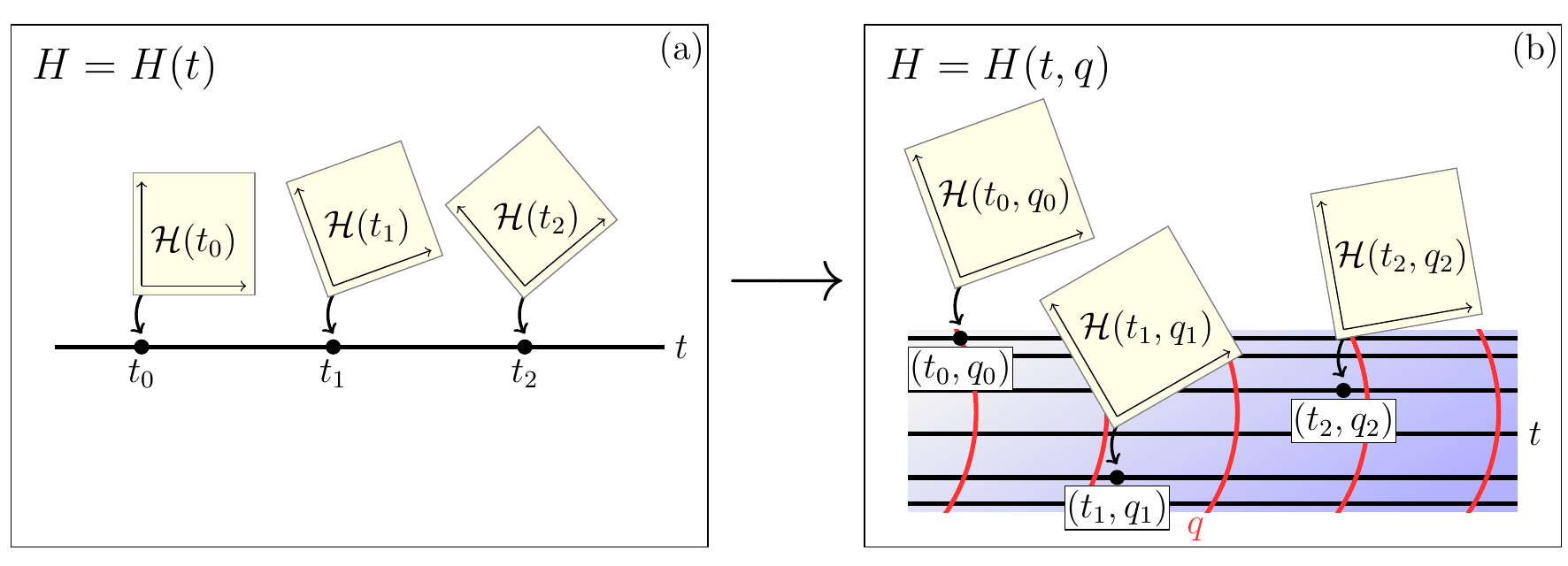}
			\end{center}
			\caption{An illustration of an emergent dimension. (a) A Hilbert space bundle, where the $t$-dimension (time dimension) forms a base space and a corresponding Hilbert space $\mathcal{H}(t)$ is equipped with a metric operator $G(t)$ at any time $t$. (b) If the Hamiltonian depends on a continuous parameter $q$; then a dimension emerges in the base space. At any base space $(t, q)$, the corresponding Hilbert space $\mathcal{H}(t, q)$ is equipped with $G(t, q)$.\label{Fig:EmergentDimension}}
		\end{figure*}

		In this study, we find that if the system has some tunable continuous parameters, these open new dimensions to the evolution space, or a base space in the fiber-bundle terminology, in addition to time $t$ (see Fig.~\ref{Fig:EmergentDimension}). We can, then, use this formalism to find how the states and the geometry of the Hilbert space vary according to some continuous physical parameters.

		For example, if an electron is placed in a magnetic field $H = H (B, \theta, \phi)$, where $B$ is the magnitude and $\theta$ and $\phi$ represent the direction of the magnetic field, the Hilbert space metric $G$ should depend on these parameters, i.e., $G = G(t, B, \theta, \phi)$ (strictly speaking, it can depend on the order of tuning these parameters). With careful examinations, we show that these parameters indeed carry the properties of coordinates in the evolution space.

		Nevertheless, since the original Hilbert space bundle is not unique but subject to a gauge transformation (a change of a basis)~\cite{Ju2019}, the gauge freedom is inherited by the induced evolution of the new dimensions. To better understand the geometry of the Hilbert space bundle, we calculate the components of the local curvature two-form~\cite{Collier2021, Needham2021, Emam2021}, an analog of the Riemann curvature tensor in GR or, more accurately, the field strength in the Yang-Mills theory. Despite that the Hilbert space bundle metrics are nontrivial in non-Hermitian quantum systems, we show that the curvature of the Hilbert space bundle is zero for any closed quantum system.

		Since the parallel transports found here are given for arbitrary quantum states, the Berry connections~\cite{Sakurai2017, Mehri2008} can be obtained by ``projecting'' the evolution equations onto a specific subspace. We also show that the Berry curvatures~\cite{Nakahara2003, Xiao2010} are indeed gauge invariant under adiabatic evolution.

		The fidelity susceptibility~\cite{Wang2015, Tzeng2021, Tu2022} is closely related to the emergent dimension evolution generator. Here we show that the fidelity susceptibility of a given eigenstate is the standard deviation squared of the evolution generator of the state. Hence, the emergent parallel transport can be used as an additional analytical tool to study the fidelity susceptibility.

	\section{Framework}

		\subsection{A new evolution dimension}
			Here we show that if a Hamiltonian, playing a similar role of the Christoffel symbol in the fiber bundle (i.e., a Christoffel-symbol-like operator), does not only depend on time $t$ (can also be time-independent) but also depends on a continuous parameter $q$, i.e., $H = H(t, q)$, the parameter becomes (or induces) an additional base-space dimension of the bundle (see Appendix~\ref{Appendix:HilbertSpaceBundle} for more detail). To show this, we first begin with the Schr\"{o}dinger equation, i.e.,
			\begin{align}
				\nabla_t \ket{ \psi } = ( \partial_t + i H ) \ket{ \psi } = 0, \label{SchrodingerEquation}
			\end{align}
			where $\nabla_t$ is a covariant derivative or a connection in a Hilbert space bundle (see Appendix~\ref{Appendix:ConventionalQM}). Although the original states only depend on time $t$, since the governing equation varies with $q$, the dynamics of the states also varies with different choice of $q$; in other words, the states should also depend on the parameter $q$, because the Hamiltonian depends on $q$. 

			Treating $\nabla_t$ as a covariant derivative (or connection) along the $t$-direction~\cite{Ju2019} leads to a self-consistent QM if the dual state of $\Ket{\psi } = \ket{\psi }$ becomes 
			\begin{align}
				\Bra{\psi } \equiv \bra{\psi } G ,
			\end{align}
			where $G $ is the metric operator of the Hilbert space bundle that satisfies $G = G^\dagger $ (so that $\Braket{\psi_A}{\psi_B} = \overline{\Braket{\psi_B}{\psi_A}}$); $G $ is positive definite (i.e., $\Braket{\psi}{\psi} \geq 0$); and
			\begin{align}
				0 = \nabla_t G = & ~ \partial_t G - i G H + i H^\dagger G . \label{MetricEq}
			\end{align}
			Analogous to the states, because the Hamilontian $H$ in Eq.~\eqref{MetricEq} varies with $q$, the metric $G$ also varies with $q$. 

			Note that additional (not-yet-known) constraints on $G$ are needed in infinite-dimensional fiber cases to ensure the finiteness of the inner products. Nevertheless, it should not affect the discussion in this work because the discussion of framework in this study is formal although based on the normalization of physical states.

			 As mentioned before, $G = G^\dagger $ by construction; therefore, we can always find an operator $\widetilde{K}$ such that the $q$-derivative of $G$ is
			\begin{align}
				\partial_q G = i G \widetilde{K} - i \widetilde{K}^\dagger G. \label{PreConnection}
			\end{align}
			The state evolution in the $q$-direction can be constructed using Eq.~\eqref{PreConnection} together with one additional assumption, namely, physical states, which are normalized, remain physical when propagating from point $(t_\text{i}, q_\text{i})$ to point $(t_\text{f}, q_\text{f})$ along path $p(s)$ [${p: s \mapsto (t(s), q(s))}$, where $s \in [0, 1]$, $(t(0), q(0)) = (t_\text{i}, q_\text{i})$, and $(t(1), q(1)) = (t_\text{f}, q_\text{f})$]; i.e.,
			\begin{align}
				\Braket{\psi (p(s))}{\psi (p(s))} = 1, \label{Assumption}
			\end{align}
			for any $s$. Hence, by choosing $p(s)$ with a constant time (i.e., ${p: s \mapsto (t, q(s))}$), we have
			\begin{align}
				& 0 = \frac{d}{ds} \Braket{\psi}{\psi} = \frac{dq}{ds} \partial_q \Braket{\psi}{\psi}\\
				& \Rightarrow 0 = \partial_q \Braket{\psi }{\psi } = \partial_q \bra{\psi } G \ket{\psi }\label{PhysicalConstraint}\\
				& \Rightarrow \partial_q \ket{\psi } = - i \widetilde{K} \ket{\psi } + \ket{\zeta }, \label{AlmostConnection}
			\end{align}
			where $\ket{\zeta }$ is a state satisfying
			\begin{align}
				\Braket{\zeta }{\psi } + \Braket{\psi }{\zeta } = 0.
			\end{align}
			However, since $\ket{\psi }$ is arbitrary, the state $\ket{\zeta }$ has to be
			\begin{align}
				\ket{\zeta } = - i \widetilde{A} \ket{\psi },
			\end{align}
			where $\widetilde{A} $ is an operator that satisfies
			\begin{align}
				G \widetilde{A} = \widetilde{A}^\dagger G .
			\end{align}

			We can, therefore, define an operator
			\begin{align}
				K = \widetilde{K} + \widetilde{A} ,
			\end{align}
			so that Eqs.~(\ref{PreConnection},~\ref{AlmostConnection}) become
			\begin{align}
				\partial_q G & = i G \widetilde{K} - i \widetilde{K}^\dagger G = i G K - i K^\dagger G . \label{MetricQEvolution}
			\end{align}
			and
			\begin{align}
				\partial_q \ket{\psi } = - i \widetilde{K} \ket{\psi } - i \widetilde{A} \ket{\psi } = - i K \ket{\psi }. \label{StateQEvolution}
			\end{align}
			Our detailed derivations, from Eq.~\eqref{MetricEq} to Eq.~\eqref{MetricQEvolution}, can be found in Appendix~\ref{Appendix:DetailedDerivation}.

			We can, therefore, naturally treat Eq.~\eqref{StateQEvolution} as
			\begin{align}
				0 = \nabla_q \ket{\psi } = ( \partial_q + i K ) \ket{\psi }, \label{StateCovariantDerivative}
			\end{align}
			where $\nabla_q$ is the induced covariant derivative (or connection) along the $q$-direction in the Hilbert space bundle with the base space extended to $M^\text{E} = \mathbb{R} \times Q$, where $Q$ is the parameter space of $q$ (see Appendix~\ref{Appendix:ExtendedBaseSpace}), and $K $ is the evolution generator, which also plays the role of the Christoffel symbol. Therefore, the state evolution in the $q$-direction is also a parallel transport. Naturally, $G$ is the metric of the Hilbert space with the base space $M^\text{E}$, and Eq.~\eqref{MetricQEvolution} suggests
			\begin{align}
				\begin{split}
					0 = \nabla_q G = \partial_q G - i G K + i K^\dagger G ,
				\end{split} \label{MetricCovariantDerivative}
			\end{align}
			which implies that $\nabla_q$ is a metric-compatible connection and $G $ is, indeed, covariantly constant.

			From Eq.~\eqref{StateCovariantDerivative} and \eqref{MetricCovariantDerivative}, it is clear that we can enlarge the base space of the Hilbert space bundle from $\mathbb{R}$ (time dimension) to $\mathbb{R} \times Q$, where $Q$ is the one-dimensional manifold describing the space of parameter $q$.

			Note that the metric operator in Eq.~\eqref{MetricEq} is not uniquely determined. The non-uniqueness is, in fact, a manifestation of its gauge freedom~\cite{Ju2019}. How the evolution generator inherits the gauge freedom is shortly detailed below.

			The governing equations of states and the metric in the $t$- and $q$-directions are summarized in Table \ref{ComparingTable}.

		\subsection{Local curvature and evolution generator}
			Although the $q$-evolution generator $K$ is still undetermined, the assumption in Eq.~\eqref{Assumption} already determines the local curvature. To be more specific, the curvature two-form is
			\begin{align}
				\mathcal{F} = \frac{1}{2} ( F_{tq} dt \wedge dq + F_{qt} dq \wedge dt ), \label{CurvatureTwoForm1}
			\end{align}
			where
			\begin{align}
				i F_{t q} \ket{\psi } \equiv \left[ \nabla_t, \nabla_q \right] \ket{\psi }, \label{CurvatureComponent}
			\end{align}
			and $F_{t q} = - F_{q t} $. Unlike the case when the base-space is one-dimensional, the curvature two-form is not identically zero~\cite{Nakahara2003, Nash2011}. Nevertheless, Eq.~\eqref{CurvatureComponent} directly leads to
			\begin{align}
				i F_{t q} \ket{\psi } & = \nabla_t \nabla_q \ket{\psi } - \nabla_q \nabla_t \ket{\psi } = \nabla_t (\nabla_q \ket{\psi }) - \nabla_q (\nabla_t \ket{\psi }) = 0.\label{Ftq}
			\end{align}
			The last equality in Eq.~\eqref{Ftq} comes from Eqs.~\eqref{SchrodingerEquation} and \eqref{StateCovariantDerivative}. Therefore, the Schr\"{o}dinger equation and the assumption that physical states remain physical leads to a vanishing local curvature, i.e., $\mathcal{F} = 0$, despite the choice of the base space $Q$. Here we want to emphasize that local flatness does not mean that the Hilbert space bundle is necessarily physically uninteresting. Many interesting fiber bundles (including the flat Mobius strip and the Klein bottle, for example) with nontrivial topologies are locally flat.

			Note that the $q$-evolution generator $K $ can also be determined from $F_{t q} = 0$ up to a gauge choice. To be more specific, since
			\begin{align}
				i F_{t q} = \left[ \nabla_t, \nabla_q \right] = i \partial_t K - i \partial_q H + \left[ K , H \right],
			\end{align}
			vanishing local curvature leads to
			\begin{align}
				\partial_t K = i \left[ K , H \right] + \partial_q H . \label{KEquation}
			\end{align}

			\begin{table}[h]
				\renewcommand*{\arraystretch}{1.8}
				\begin{center}
					\begin{tabular}{| c | c | c |}
						\hline
						& Evolution in $t$-direction & Evolution in $q$-direction\\
						\hline
						State & $\partial_t \ket{\psi} = - i H \ket{\psi}$ & \begin{tabular}{l}$\partial_q \ket{\psi} = - i K \ket{\psi}$,\\$\partial_t K = i \left[K, H \right] + \partial_q H$\end{tabular}\\
						\hline
						Metric & $\partial_t G = i \left(G H - H^\dagger G\right)$ & $\partial_q G = i \left(G K - K^\dagger G\right)$\\
						\hline
					\end{tabular}
				\end{center}
				\caption{Comparison of the equations of motion along the $t$- and $q$-directions. The operator $K$ in the table can be computed using Eq.~\eqref{KEquation}.}
				\label{ComparingTable}
			\end{table}

		\subsection{Gauge degrees of freedom}
			It is obvious that $K $ is not unique because Eq.~\eqref{KEquation} is a differential equation. Thus,
			\begin{align}
				K' = K + \Delta K ,
			\end{align}
			with $\Delta K $ being a homogeneous solution of Eq.~\eqref{KEquation}, i.e.,
			\begin{align}
				\partial_t \Delta K = i \left[ \Delta K , H \right]
			\end{align}
			also provides a valid connection.

			As a matter of fact, this freedom of choosing $K $ originates from the gauge symmetry of the metric $G $~\cite{Ju2019}, i.e., the non-uniqueness of the metric operator $G $. More specifically, $G $ {is transformed to}
			\begin{align}
				G' = T^\dagger G T 
			\end{align}
			is a valid metric if the operator $T $ is invertible and satisfies
			\begin{align}
				\partial_t T + i [H , T ] = 0. \label{GaugeTransformation}
			\end{align}

			To show that the freedom of choosing $K $ results from the gauge choice of $G $, we start {from} $G' $, {for} $T \in$~GL($n, \mathbbm{C}$), which satisfies Eq.~\eqref{GaugeTransformation}. A direct calculation shows
			\begin{align}
				\partial_q G' = i G K' - i K^{\prime \dagger} G ,
			\end{align}
			where
			\begin{align}
				K' = T^{-1} K T - i T^{-1} \partial_q T .
			\end{align}
			Taking the $q$-derivative of $K' $ gives
			\begin{align}
				\partial_q K' = i [K' , H ] + \partial_q H .
			\end{align}
			Therefore, $\Delta K = K' - K $ is indeed a homogenous solution of Eq.~\eqref{KEquation}, which has the same degrees of freedom as that induced from the gauge degrees of freedom in $G $. Hence, different choices of $K $ merely represent different choices of the gauge of $G $.

			It is worth mentioning that $K $ can always be chosen Hermitian when $H = H^\dagger $, since $K^\dagger $ also satisfies Eq.~\eqref{KEquation}. From Eq.~\eqref{MetricCovariantDerivative}, when $G = \mathbbm{1}$ and $K = K^\dagger $, the metric is always the identity because $\partial_q G = 0$. In other words, this formalism is fully compatible with conventional QM for Hermitian quantum systems.

		\subsection{Multiple dimensions}
			 Although the previous discussion focuses on Hamiltonians with only one parameter, besides time $t$, the method can also be applied to Hamiltonians with multiple parameters, $H = H(t, q_1, q_{2}, \cdots, q_n) = H(t, \lbrace q \rbrace)$, where $\lbrace q \rbrace$ is short for $q_1$, $q_2$, $\cdots$, $q_n$. That is, every parameter of the Hamiltonian represents a dimension in the base space. Analogous to the discussion for the single parameter case, the base space $\mathbb{R}$ (i.e., the time dimension) is now extended to $\mathbb{R} \times Q^n$, where $Q^n$ is a $n$-dimensional manifold of the parameter space (see Appendix~\ref{Appendix:ExtendedBaseSpace} for more details). Applying the same procedure on every $q_i$, we obtain
			\begin{align}
				\nabla_{i} \ket{\psi } = (\partial_{i} + i K_{i} ) \ket{\psi } = 0, \label{StateQiEvolution}
			\end{align}
			and
			\begin{align}
				0 = \nabla_{i} G = \partial_{i} G - i G K_{i} + i K_{i}^\dagger G , \label{MetricQiEvolution}
			\end{align}
			where $\nabla_{i}$ is the covariant derivative along the $q_i$-direction and $\partial_i$ is short for $\partial / \partial q_i$. Note that $K_{i} $ is governed by
			\begin{align}
				0 = F_{t i} = \partial_t K_{i} - \partial_{i} H - i \left[ K_{i} , H \right]. \label{MultiKEquation1}
			\end{align}
			Moreover, Eq.~\eqref{StateQiEvolution} implies
			\begin{align}
				& i F_{ij} \ket{\psi } = \left[ \nabla_{i}, \nabla_{j} \right] \ket{\psi } = 0\\
				\Rightarrow ~ & 0 = F_{ij} = \partial_{i} K_{j} - \partial_{j} K_{i} - i \left[ K_{j} , K_{i} \right]. \label{MultiKEquation2}
			\end{align}
			Therefore, the Hilbert space bundle curvature remains vanishing even if more dimensions are added to the generalized evolution space, because all the components of the curvature two-form,
			\begin{align}
				\mathcal{F} = & ~ \frac{1}{2} \bigg(\sum_iF_{ti} dt \wedge dq_i + \sum_i F_{it} dq_i \wedge dt + \sum_{ij}F_{ij} dq_i \wedge dq_j \bigg), \label{CurvatureTwoForm2}
			\end{align}
			are zero.

		\subsection{Brief summary}
			In conventional QM, quantum states are defined in a Hilbert space $\mathcal{H}$. To include the dynamics of the states, the space can be generalized to a Hilbert space bundle, where the fiber is the Hilbert space $\mathcal{H}$ and the base space is the time-dimension, t{. T}hat is, $M = \mathbb{R}$, where the time evolution of the states can be described as a parallel transport along $M$, as described by the Schr\"{o}dinger equation [Eq.~\eqref{SchrodingerEquation}].

			However, if the system Hamiltonian depends on $n$ continuous parameters (the $q_i$'s), then the base space is extended to $M^\text{E} = \mathbb{R} \times Q^n$, where $\mathbb{R}$ is the time dimension and $Q^n$ is an $n$-dimensional manifold that describes the parameter space. Like in the evolution in the time dimension, the evolutions in the parameter space (the sub-manifold $Q^n$) can also be interpreted as the parallel transports according to Eq.~\eqref{StateQiEvolution}.

			Despite the fact that the extended manifold can be nontrivial, every components of the curvature two-form vanishes [Eq.~\eqref{MultiKEquation1} and \eqref{MultiKEquation2}]; hence the Hilbert space bundle is locally flat. Nevertheless, we should emphasize that a locally flat space does not imply that the global behavior is trivial.

			Note that even though we have assumed $G \geq 0$ so that the fiber space is a Hilbert space (i.e., $\Braket{\psi}{\psi} = \bra{\psi} G \ket{\psi} > 0$ if $\ket{\psi} \neq 0$), this constraint can be relaxed to potentially extend the discussion from Hilbert space to other spaces (e.g., the spaces having spurious states or other non-physical states $\Braket{\psi}{\psi} = \bra{\psi} G \ket{\psi} \leq 0$)~\cite{Polchinski1998, Becker2006}.

	\section{Applications}

		\subsection{Gauge fixing for time-independent Hamiltonians}
			Although all $K$s can be derived from Eq.~\eqref{KEquation} or Eqs.~\eqref{MultiKEquation1} and \eqref{MultiKEquation2}, finding the most general solution and then fix the gauge is not the most desirable procedure. There are many examples where good choices of gauge-fixing conditions lead to significant results~\cite{Stefano2019, Garziano2020, Settineri2021, Savasta2021, Salmon2022}. Here we provide a gauge-fixing condition for time-independent systems that could reduce the complexity of calculations.

			We first discuss the single-parameter case, which can be easily transferred to the multiparameter case. For a time-independent Hamiltonian, i.e., $H = H (q)$, we apply
			\begin{align}
				\left[ \partial_t K , H \right] = 0, \label{AdiabaticGaugeCondition}
			\end{align}
			as a gauge-fixing condition. This gauge-fixing condition guarantees that the state evolution is adiabatic, namely,
			\begin{align}
				H \ket{\psi} = h \ket{\psi}.
			\end{align}
			Moreover, the eigenvalues of $\partial_t K$ are the $q$-derivative of the eigenvalues of $H $. Detailed derivations can be found in Appendix~\ref{Appendix:AdiabaticGauge}.

			A direct consequence of this gauge-fixing condition is that it turns Eq.~\eqref{KEquation}, which is a differential equation, into an algebraic equation.

			Using $\partial_t H = 0$, by taking a time derivative on both sides of Eq.~\eqref{KEquation} together with the gauge fixing condition in Eq.~\eqref{AdiabaticGaugeCondition}, we arrive at
			\begin{align}
				\partial_t^2 K = 0 \quad \Rightarrow \quad K = t K^{(1)} + K^{(0)}, \label{LinearInTime}
			\end{align}
			where $K^{(1)}$ and $K^{(0)}$ are both time-independent operators.

			Substituting the $K$ in Eqs.~\eqref{KEquation} and \eqref{AdiabaticGaugeCondition} with Eq.~\eqref{LinearInTime}, we find
			\begin{align}
				& K^{(1)} = i \left[ K^{(0)} , H \right] + \partial_q H , \label{AlgebraicEq1}\\
				& [K^{(1)} , H ] = 0, \label{AlgebraicEq2}
			\end{align}
			where the last equation comes from the fact that $K^{(1)} $ and $H $ share the same eigenstates. Detailed derivations and an example can be found in Appendices~\ref{Appendix:AlgebraicEquations} and \ref{Appendix:FidelitySusceptibility}.

			The $K^{(1)} $ can be determined algebraically from the equations above, while the $K^{(0)} $ is almost fixed up to a time-independent gauge freedom, $\Delta K $, which satisfies $\left[ \Delta K , H \right] = 0$. That is, there are some residual gauge degrees of freedom using the adiabatic gauge fixing condition. These degrees of freedom are the manifestations of two well-known properties, namely, the freedom of multiplying an eigenstate with a non-zero constant and that of the ``rotation'' between the eigenstates of the same eigenvalue. 

			The above discussion can also be generalized to multiparameter systems. The adiabatic gauge-fixing conditions are
			\begin{align}
				\begin{array}{c}
					\displaystyle \left[ \partial_t K_{i} , H \right] = 0\vspace{0.1cm},\\
					\displaystyle \left[ \partial_t K_{i} , \partial_t K_{j} \right] = 0,
				\end{array} \label{AdiabaticGaugeConditions}
			\end{align}
			where the second vanishing commutation relation comes from that $H $ and $\partial_t K_{i} $ share the same eigenstates. A direct calculation shows that these equations render
			\begin{align}
				K_{i} = t K_{i}^{(1)} + K_{i}^{(0)} , \label{LinearInTimeMultiple}
			\end{align}
			where
			\begin{align}
				& K_{i}^{(1)} = i \left[ K_{i}^{(0)} , H \right] + \partial_{i} H ,\\
				& [K_{i}^{(1)} , H ] = 0,\\
				& [K_{i}^{(1)} , K_{j}^{(1)} ] = 0. \label{KKCommute}
			\end{align}
			Moreover, $K^{(0)} $ and $K^{(1)} $ are further related to each other through Eq.~\eqref{MultiKEquation2}, which leads to
			\begin{align}
				& \partial_{i} K_{j}^{(1)} - \partial_{j} K_{i}^{(1)} = i \left[ K_{j}^{(0)} , K_{i}^{(1)} \right] - i \left[ K_{i}^{(0)} , K_{j}^{(1)} \right], \label{tTerm}\\
				& \partial_{i} K_{j}^{(0)} - \partial_{j} K_{i}^{(0)} = i \left[ K_{j}^{(0)} , K_{i}^{(0)} \right]. \label{ConstantTerm}
			\end{align}
			Equation~\eqref{tTerm} comes from the $t$ term in Eq.~\eqref{KKCommute} and Eq.~\eqref{ConstantTerm} comes from the constant term, where $t^2$ vanishes automatically due to Eq.~\eqref{KKCommute}, except when $H (\lbrace q \rbrace)$ is at an excpetional point (EP), where the operators $K$ are already singular as discussed in Appendix~\ref{Appendix:FidelitySusceptibility}. A multiple parameter system example can be found in Appendix~\ref{Appendix:BerryCuravture}.

			Some physical quantities are demonstrated below showing their relations to the adiabatic gauge and its advantage.

		\subsection{Berry connections and curvature}
			As an example, we show that the above-mentioned geometric understanding can also be applied to Hermitian systems. Here, we focus on $H (\lbrace q \rbrace) = H^\dagger (\lbrace q \rbrace)$ and $G = \mathbbm{1}$, so that $\Ket{\psi } = \ket{\psi }$ and $\Bra{\psi } = \bra{\psi }$.

			It is known that the Berry connections are the connections of a specific eigenstate of the Hamiltonian, and different eigenstates generally have different connections. However, the Hilbert space bundle connections discussed in this paper are not limited to any states, but are general properties of the whole Hilbert space bundle. Therefore, we can reduce $K_{i} $ to the Berry potentials, $\mathcal{A}_{i}^n $, through simple projections to the eigenstates, i.e.,
			\begin{align}
				\mathcal{A}_{i}^n & = i \bra{\psi_n } \partial_{i} \ket{\psi_n } = \bra{\psi_n } K_{i} \ket{\psi_n },
			\end{align}
			where $\braket{\psi_m }{\psi_n } = \delta_{mn}$, and $\mathcal{A}_{i}^n $ is the Berry potential of the $n^\text{}$th eigenstate along the $q_i$-direction in an adiabatic process~\cite{Born1928}. That is to say, $K_i $ contain all the information about the Berry potentials applying within an adiabatic gauge in Eq~\eqref{AdiabaticGaugeConditions}.

			It is well known that the Berry potentials are not gauge invariant (and neither are $K_i $). Nevertheless, for a nondegenerate Hamiltonian, the Berry curvature,
			\begin{align}
				\Omega_{ij}^n = \partial_{i} \mathcal{A}_{j}^n - \partial_{j} \mathcal{A}_{i}^n ,
			\end{align}
			turns out to be time-independent and gauge invariant under the residual gauge transformation. This is indeed consistent with the standard Berry curvature property, which has been used to find some topological invariants~\cite{Berry1984, Nandy2018}. Nevertheless, this was not expected in the sense that the gauge transformation of the potentials was restricted to the eigenstate that the Berry potentials are defined on, but the Berry curvature also turned out to be gauge invariant under the whole Hilbert space gauge transformations. Moreover, the curvature of the full Hilbert space bundle is always zero, but Berry curvatures do not need to be zero.

			To show that the Berry curvature is invariant under the residual gauge transformation of $K$, we first show the relation of $\Omega_{ij}^n $ to $K $:
			\begin{align}
				\begin{split}
					\Omega_{ij}^n & = \partial_{i} \bra{\psi_n } K_{j} \ket{\psi_n } - \partial_{j} \bra{\psi_n } K_{i} \ket{\psi_n }\\
					& = i \bra{\psi_n } \left[ K_{i} , K_{j} \right] \ket{\psi_n }\\
					& = i \bra{\psi_n } \left[ K_{i}^{(0)} , K_{j}^{(0)} \right] \ket{\psi_n },
				\end{split}
			\end{align}
			where Eq.~\eqref{MultiKEquation2} was used in the derivation and the last equality is due to $\ket{\psi_n }$ being simultaneously an eigenstate of $K_{i}^{(1)} $, $K_{j}^{(1)} $, and $H$. {R}ecall that
			\begin{align}
				\ket{\psi_n (t, \lbrace q \rbrace)} = {\exp [- i t h_n (\lbrace q \rbrace)]}\ket{\psi_n (0, \lbrace q \rbrace)},
			\end{align}
			where $h_n (\lbrace q \rbrace)$ is the corresponding eigenvalue of the eigenstate $\ket{\psi_n (t, \lbrace q \rbrace)}$. Thus, the $\Omega_{ij}^n$ is, in fact, time-independent; i.e., $\partial_t \Omega_{ij}^n = 0$.

			We next discuss the Berry curvature under the residual gauge transformation of $K $. Let
			\begin{align}
				K'_{i} = K_{i} + \Delta K_{i} ,
			\end{align}
			where $\Delta K_{i} $ is the {residual gauge transformation} (the time-independence of the residual gauge transformation is explained previously). The Berry curvature after applying the {residual gauge transformation} {becomes}
			\begin{align}
				\begin{split}
					\Omega_{i j}^{\prime n} & = i \bra{\psi_n } \left[ K_{i}' , K_{j}' \right] \ket{\psi_n }\\
					& = i \bra{\psi_n } \left( \left[ K_{i} + \Delta K_{i} , K_{j} + \Delta K_{j} \right] \right) \ket{\psi_n }\\
					& = i \bra{\psi_n } \left[ K_{i} , K_{j} \right] \ket{\psi_n } = \Omega_{ij}^n ,
				\end{split}
			\end{align}
			where the third equality comes from the fact that $\left[ \Delta K_{i} , H \right] = 0$ and $\ket{\psi_n }$ being an eigenstate of $H $. Hence, the Berry curvature is indeed invariant under the {residual gauge transformation} of $K$. An example of acquiring Berry curvature using the generator $K$s can be found in Appendix~\ref{Appendix:BerryCuravture}.

		\subsection{Fidelity susceptibility}
			It is well established that the fidelity between the eigenstates of similar Hamiltonians, $H(q)$ and $H(q + \epsilon)$, can be used to detect phase transitions~\cite{Gu2010}. To be more specific, the divergence of the fidelity susceptibility, to be defined shortly, is a sign of a phase transition. Here, we provide a way to look at the fidelity susceptibility from a different aspect.

			For states $\ket{\psi }$ and $\ket{\phi }$ in Hermitian systems, the fidelity between them is defined to be
			\begin{align}
				\mathcal{F}_\text{H}\left( \ket{\psi } , \ket{\varphi }\right) = \left| \braket{\psi }{\varphi } \right|^2.
			\end{align}
			That is, the fidelity between $\ket{\psi_n(t, q)}$ and $\ket{\psi_n(t, q + \epsilon)}$ for small $\epsilon$, the $n$th normalized eigenstates of Hamiltonians $H(q)$ and $H(q + \epsilon)$, can be expanded as
			\begin{align}
				\mathcal{F}_\text{H}\left( \ket{\psi_n(t, q)} , \ket{\psi_n(t, q + \epsilon)} \right) = 1 - \epsilon^2 \chi_n(q) + \mathcal{O}\left(\epsilon^3\right),
			\end{align}
			where $\chi_n(q)$ is called the fidelity susceptibility. Its time-independence is shown below.

			When the ground state fidelity susceptibility diverges at some system parameters, 
			\begin{align}
				\lim_{q \rightarrow q_\text{\tiny PT}}\chi_0(q) \rightarrow \infty,
			\end{align}
			the system exhibits a phase transition at the parameter ($q = q_\text{\tiny PT}$).

			Taking the geometries of the Hilbert spaces into account, the fidelity has been generalized~\cite{Tzeng2021} to
			\begin{align}
				\mathcal{F}_\text{G}\left( \ket{\psi_n(t, q)} , \ket{\psi_n(t, q + \epsilon)} \right) = \Braket{\psi_n(t, q)}{\psi_n(t, q + \epsilon)}\Braket{\psi_n(t, q + \epsilon)}{\psi_n(t, q)}, \label{GenerlizedFidelity}
			\end{align}
			where
			\begin{align}
				\Bra{\psi_n(t, q)} = \bra{\psi_n(t, q)} G(t, q)
			\end{align}
			and $\Bra{\psi_n(t, q + \epsilon)} = \bra{\psi_n(t, q + \epsilon)} G(t, q + \epsilon)$, with $G(t, q)$ and $G(t, q + \epsilon)$ being the metric for $H(q)$ and $H(q + \epsilon)$, respectively. Moreover, we set $\Braket{\psi_m(t, q)}{\psi_n(t, q)} = \delta_{mn}$.

			Note that the adiabatic gauge fixing condition in Eq.~\eqref{AdiabaticGaugeCondition} is applied here so that every $\Ket{\psi_n(t, q)}$ is an eigenstate of $H(q)$ for all $q$.

			Expanding Eq.~\eqref{GenerlizedFidelity} in $\epsilon$, we find
			\begin{align}
				\mathcal{F}_\text{G} = & 1 - \epsilon^2 \Big[ \Bra{\psi_n(t, q)} K^2(t, q) \Ket{\psi_n(t, q)} - \Bra{\psi_n(t, q)} K(t, q) \Ket{\psi_n(t, q)}^2 \Big] + \mathcal{O}\left( \epsilon^3 \right).
			\end{align}
			Therefore, the generalized fidelity susceptibility becomes
			\begin{align}
				\chi_n (q) = & ~ \Bra{\psi_n(t, q)} K^2(t, q) \Ket{\psi_n(t, q)} - \Bra{\psi_n(t, q)} K(t, q) \Ket{\psi_n(t, q)}^2. \label{Susceptibility}
			\end{align}
			Since $H (q)$ is time-independent, according to Eq.~\eqref{LinearInTime},
			\begin{align}
				K (t, q) = t K^{(1)} (q) + K^{(0)} (q){,}
			\end{align}
			then Eq.~\eqref{Susceptibility} becomes
			\begin{align}
				\chi_n (q) = & ~ \Bra{\psi_n (t, q)} \left( K^{(0)} (q) \right)^2 \Ket{\psi_n (t, q)} - \Bra{\psi_n (t, q)} K^{(0)} (q) \Ket{\psi_n (t, q)}^2\\
				= & ~ \Bra{\psi_n (0, q)} \left( K^{(0)} (q) \right)^2 \Ket{\psi_n (0, q)} - \Bra{\psi_n (0, q)} K^{(0)} (q) \Ket{\psi_n (0, q)}^2,
			\end{align}
			where $K^{(1)} (q)$ does not contribute because $[K^{(1)} (q), H (q)] = 0$ and, hence, is time-independent. Moreover, $\chi_i (q)$ is time-independent and invariant under the residual gauge freedom since the difference between gauges also commutes with the Hamiltonian, i.e., $[ \Delta K (q), H (q)] = 0$.

			Note that if $H\left(q\right)$ is non-diagonalizable at $q = q_\text{\tiny EP}$, i.e., at an EP~\cite{Kato1976, Heiss2004, Ozdemir2019}, then Eqs.~\eqref{AlgebraicEq1} and \eqref{AlgebraicEq2} do not have a solution in general. Hence, $K (t, q)$, in general, becomes singular at $q = q_\text{\tiny EP}$ and explains why the fidelity susceptibility tends to diverge at the EPs. An example can be found in Appendix~\ref{Appendix:FidelitySusceptibility}.

			Moreover, it is well-known that the fidelity susceptibility generally diverges at the critical points of quantum phase transitions~\cite{Gu2010}. From Eq.~\eqref{Susceptibility}, we can deduce that the evolution generator can also be singular at the critical point because a divergent fidelity susceptibility is a necessary condition for the evolution generator to be singular. Hence, besides the conventional techniques, we provide an additional method, namely, the compatibility condition of Eqs.~\eqref{KEquation} and \eqref{AdiabaticGaugeCondition}, to study the fidelity susceptibility.

	\section{Conclusions}

		This study shows that not only closed non-Hermitian quantum systems can benefit from the geometrical treatment of QM. By treating quantum mechanics geometrically, we derived some additional Schr\"{o}dinger-like equations that govern the evolution of states and the Hilbert space bundle metric in parameter space manifold. These equations, inspired by non-Hermitian QM, can also be applied to Hermitian quantum systems and provide some useful physical quantities, such as the Berry connection, Berry curvature, and fidelity susceptibility. Note that a similar study, using emergent geometrical properties of the adiabatic process, can be found in \cite{Rattacaso2020}.

		Despite that the Berry curvature (i.e., the ``curvature'' of a certain subspace) can be non-zero, we find that the full Hilbert space bundle curvature of any closed quantum system always vanishes. Here we emphasize that flat Hilbert space bundles can still be interesting.

		Moreover, this geometric treatment of QM does not only provide an additional tool to analyze phase transitions (via the relation between the emergent dimension evolution generator and the fidelity susceptibility), but it can also potentially lead to some deeper understanding of unresolved physics problems, the discovery of new physical phenomena, or new topological classification methods of quantum systems, e.g., providing a new analytical method for calculating a topological entanglement entropy, understanding the behavior of quantum systems crossing a non-Hermitian exceptional point, finding an event-horizon-like quantum behavior, or defining new topological phases.

	\begin{acknowledgments}
		 C.Y.J. would like to thank Chia-Min Chung and Yu-Chin Tzeng for fruitful discussions. C.Y.J. is partially supported by the National Science and Technology Council (NSTC) through Grant No. NSTC 112-2112-M-110-013-MY3 and the Ministry of Science and Technology (MOST) through Grant No. 111-2112-M-110-007-MY2, and the National Center for Theoretical Sciences (NCTS). A.M. is supported by the Polish National Science Centre (NCN) under the Maestro Grant No. DEC-2019/34/A/ST2/00081. Y.N.C acknowledges the support of the U.S. Army Research Office (ARO Grant No. W911NF-19-1-0081) and the NCTS. G.Y.C. is partially supported by the NCTS, the MOST through Grant No. MOST 110-2112-M-005-002, and NSTC through Grant No. 112-2112-M-005-006. Both Y.N.C. and G.Y.C are partially supported by the NSTC through Grant No. 111-2123-M-006-001 and 112-2123-M-006-001. F.N. is supported in part by: Nippon Telegraph and Telephone Corporation (NTT) Research, the Japan Science and Technology Agency (JST) [via the Quantum Leap Flagship Program (Q-LEAP) and the Moonshot R\&D Grant Number JPMJMS2061], the Asian Office of Aerospace Research and Development (AOARD) (via Grant No. FA2386-20-1-4069), and the Office of Naval Research Global (ONR) (via Grant No. N62909-23-1-2074).
	\end{acknowledgments}

	\vspace*{-0.2cm}\section*{Appendices}
		\begin{appendix}
			\section{Comparison between metricized quantum mechanics and general relativity\label{Appendix:MQMandGR}}

				In metricized quantum mechanics, the Hilbert space is equipped with a metric so that the relation between a vector (state) and its dual vector (dual state) is not merely a complex conjugation but can also be subject to a linear transformation, just like the relation between a vector and the corresponding dual vector in GR. In other words, the dual state ($\Bra{\psi}$) of a state ($\Ket{\psi} = \ket{\psi}$) is not the standard bra vector ($\bra{\psi}$) but needs to be linearly transformed by the metric operator ($\Bra{\psi} = \bra{\psi} G$, where $G$ is the metric operator).

				Nevertheless, the vectors in GR live in the tangent space of the manifold, while the vectors in QM live in a Hilbert space that is not related to the manifold in QM. Although they seem different, they both fall into the category of fiber bundles. The base space of the bundle is a manifold, while the fiber of the bundle is the tangent space of the manifold in GR and a Hilbert space in QM.

				In order to determine the geometry of the bundle, it is important to know how a vector propagates from one point to another on the fiber. It is well-established that the Christoffel symbol $\Gamma^\alpha_{~\mu \beta}$ in GR [or the connection coefficients in (pseudo-)Riemannian geometry] relates the two overlapping charts in the $\mu$-direction (also in the tangent space). We can, therefore, single out the $\mu$ (of course, we can choose a gauge such that the coefficient is symmetric in $\mu$ and $\beta$ numerically if no fermions are involved~\cite{Freedman1976, Nieuwenhuizen1981}) and define a matrix as
				\begin{align}
					\Gamma_\mu = \begin{pmatrix}
						\Gamma^1_{~\mu 1} & \Gamma^1_{~\mu 2} & \cdots\\
						\Gamma^2_{~\mu 1} & \Gamma^2_{~\mu 2} & \cdots\\
						\vdots & \vdots & \ddots
					\end{pmatrix},
				\end{align}
				In metricized QM, the Hamiltonian plays the role of Christoffel symbol in the $t$-direction (or $\Gamma_t$, to be more precise) up to an imaginary number $i$. Therefore, the Schr\"{o}dinger equation becomes a parallel transport of a vector in the $t$-direction.

				This paper shows that if the Hamiltonian is a function of physical parameters $\lbrace q \rbrace$, the base space manifold can be extended to a larger one (see Appendix~\ref{Appendix:HilbertSpaceBundle}) that includes the parameters manifold so that the vectors in the Hilbert space bundle can propagate in the $q$-directions (see Table~II in the main text).

				Since the dimension of the base space manifold can be larger than 1, it is natural to find the curvature two-form of the Hilbert space bundle. Nevertheless, as discussed in the main text, the local curvature two-forms are always zero. This means that the Hilbert space bundle is locally flat, even if the parameter manifold is non-trivial.

				It is worth mentioning that local flatness is the part that is different from GR, where many interesting phenomena come from the non-trivial local curvature. Nevertheless, vanishing local curvature does not imply the geometry is uninteresting. For example, when the base space has a puncture or is of nonzero genus, additional information will be included in the system (e.g., winding numbers). A more physical example is also given in the main text, where the connection coefficients are used to determine quantum phase transitions.

			\section{Hilbert space bundles\label{Appendix:HilbertSpaceBundle}}

				This note describes the basic concepts of the Hilbert space bundles and some terminology used in this paper.

				\subsection{Hilbert space bundle for conventional quantum mechanics\label{Appendix:ConventionalQM}}

					We start with the standard Hilbert space $\mathcal{H}$ of quantum states. It is know that the quantum states are represented by vectors in the Hilbert space, and the scalar product between vectors $\ket{\phi}$ and $\ket{\psi}$ in $\mathcal{H}$ is defined by
					\begin{align}
						\braket{\phi}{\psi} = \overline{\braket{\psi}{\phi}},
					\end{align}
					where $\ket{\phi}$ and $\ket{\psi}$ are vectors in the Hilbert space and $\bra{\phi} = \ket{\phi}^\dagger$ and $\bra{\psi} = \ket{\psi}^\dagger$ are the dual vectors of $\ket{\phi}$ and $\ket{\psi}$, respectively.

							To include the time evolution of quantum states, it is natural to define the scalar product becomes
					\begin{align}
						\braket{\phi(t)}{\psi(t)} = \overline{\braket{\psi(t)}{\phi(t)}}, \label{Appendix:ConventionalHilbertSpaceInnerProduct}
					\end{align}
					at a certain time slice $t$; where $\ket{\phi(t)}$ and $\ket{\psi(t)}$) are vectors at time $t$ and $\bra{\phi(t)} = \ket{\phi(t)}^\dagger$ and $\bra{\psi(t)} = \ket{\psi(t)}^\dagger$ are the dual vectors of $\ket{\phi(t)}$ and $\ket{\psi(t)}$ at time $t$.

					There are now two different spaces or manifolds to describe a state, one being the temporal space ($M = \mathbb{R}$) and the Hilbert space at $t \in M$, namely, $\mathcal{H}(t)$. In order to describe a state, we now turn to the concept of fiber bundles.

					We start with the space of time $M$, which is called a base space. At each time slice $t$, a quantum state is a vector (or a local section) $\ket{\psi(t)}$ that is spanned in the Hilbert space $\mathcal{H} (t)$, which is called the fiber, at time $t$ endowed with the scalar product in Eq.~\eqref{Appendix:ConventionalHilbertSpaceInnerProduct}. The space that includes both base space and fiber is called a total space $\displaystyle E = \bigcup_{t \in M} \mathcal{H}(t)$, with a projection $\pi$ that extracts the time information of the elements in $E$ [e.g., $\pi(\ket{\psi(t)}) = t$], i.e.,
					\begin{align}
						\pi: E \rightarrow M.
					\end{align}
					The fiber bundle $E \stackrel{\pi}{\rightarrow} M$ is called a Hilbert space bundle.

					For a quantum system described by the Hamiltonian $H(t)$ (the time dependence is to make the discussion more general), the time evolution of the quantum states is governed by the Schr\"{o}dinger equation, namely,
					\begin{align}
						i \partial_t \ket{\psi(t)} = H(t) \ket{\psi(t)}. \label{Appendix:SchroedingerEq}
					\end{align}
					By defining a covariant derivative (or connection) $\nabla_t$ on the vector $\ket{\psi(t)}$ as
					\begin{align}
						\nabla_t \ket{\psi(t)} = \left[\partial_t + i H(t)\right] \ket{\psi(t)},
					\end{align}
					the Schr\"{o}dinger equation in Eq.~\eqref{Appendix:SchroedingerEq} becomes Eq.~\eqref{SchrodingerEquation}, namely, a parallel transport in the Hilbert space bundle $E \stackrel{\pi}{\rightarrow} M$.

				\subsection{Hilbert space bundle with a metric operator\label{Appendix:HilbertSpaceBundleWithG}}

					The bundle discussed above is a trivial bundle, i.e., the total space is just a direct product of the base space and the fiber. In a (non-)Hermitian quantum system, the inner product of states at each time slice can be different (see \cite{Ju2019} for more detail), namely,
					\begin{align}
						\Braket{\phi(t)}{\psi(t)} = \bra{\psi(t)} G(t) \ket{\psi(t)} = \overline{\Braket{\psi(t)}{\phi(t)}}, \label{Appendix:MetricizedHilbertSpaceInnerProduct}
					\end{align}
					where $\Ket{\psi(t)} = \ket{\psi(t)}$ and $\Ket{\phi(t)} = \ket{\phi(t)}$ are vectors in a Hilbert space and $\Bra{\phi(t)} = \bra{\phi(t)} G(t)$ and $\Bra{\psi(t)} = \bra{\psi(t)} G(t)$ are the dual vectors of $\Ket{\phi(t)}$ and $\Ket{\psi(t)}$, also at time $t$.

					Thus, at different time slices, the Hilbert space can be different. We, therefore, define $\mathcal{H}(t)$ to be the Hilbert space equipped with the scalar inner product defined in Eq.~\eqref{Appendix:MetricizedHilbertSpaceInnerProduct}. In this case, the total space of the bundle is, roughly speaking, a collection of all the Hilbert spaces, i.e.,
					\begin{align}
						E = \bigcup_{t \in M} \mathcal{H}(t),\label{Appendix:MetricE}
					\end{align}
					with a projection that picks out the time slice of the Hilbert space, i.e.,
					\begin{align}
						\pi: \bigcup_{t \in M} \mathcal{H}(t) \rightarrow M,
					\end{align}
					where $M = \mathbb{R}$ is, again, the temporal space. That is, the Hilbert space bundle in a (non-)Hermitian system is $E \stackrel{\pi}{\rightarrow} M$, where $E$ is defined in Eq.~\eqref{Appendix:MetricE}; while the time evolution of vectors (local section) is described as a parallel transport in Eq.~\eqref{SchrodingerEquation} and the metric operator $G(t)$ is governed by Eq.~\eqref{SchrodingerEquation} with $G(t) = G^\dagger(t)$ and $G(t) > 0$ (positive-definite) at any time slice $t$~\cite{Ju2019}.

				\subsection{Hilbert space bundle with extended base space\label{Appendix:ExtendedBaseSpace}}

					In the main text, we found that if the Hamiltonian depends on additional continuous parameters, i.e., $H = H\left(t, \lbrace q \rbrace\right)$, other than the parallel transport or, equivalently, the Schr\"{o}dinger equation in the time dimension [Eq.~\eqref{SchrodingerEquation}], the evolution of the vectors and metric operator on the parameter $q_i$ should also obey Eqs.~\eqref{StateQiEvolution} and \eqref{MetricQiEvolution}, respectively, with the scalar product of the Hilbert space at $(t, \lbrace q \rbrace)$ being
					\begin{align}
						\Braket{\phi(t, \lbrace q \rbrace)}{\psi(t, \lbrace q \rbrace)} = \bra{\psi(t, \lbrace q \rbrace)} G(t, \lbrace q \rbrace) \ket{\psi(t, \lbrace q \rbrace)}.\label{Appendix:QTScalarProduct}
					\end{align}

					The evolution equations for both vectors [Eq.~\eqref{StateQiEvolution}] and the metric operator [Eq.~\eqref{MetricQiEvolution}] in $q_i$'s are formally the same as the parallel transport equation for the local section and the connection-compatible condition for the fiber metric in a Hilbert space bundle. Thus,
					\begin{align}
						\bigcup_{(t, \lbrace q \rbrace) \in M^\text{E}} \mathcal{H}(t, \lbrace q \rbrace) \rightarrow M^\text{E},
					\end{align}
					where $\mathcal{H}(t, \lbrace q \rbrace)$ is a Hilbert space endowed with the scalar inner product defined in Eq.~\eqref{Appendix:QTScalarProduct} and $M^\text{E} = \mathbb{R} \times Q^n$, where $\mathbb{R}$ is the original time dimension and $Q^n$ is an $n$-dimensional manifold describing the parameter space. (Note that $Q^n$ can be nontrivial, depending on the system setup.)

					Hence, it is natural to extend the systems with continuous parameters $\lbrace q \rbrace$ from $M = \mathbb{R}$ to $M^\text{E} = \mathbb{R} \times Q^n$, because they are formally indistinguishable.

		\section{Detailed derivation of the metric induced generator\label{Appendix:DetailedDerivation}}

			Given $H = H(t, q)$, where $q$ is a parameter of the system, the metric $G $ also varies with the parameter $q$ because $G $ is related to $H $ via
			\begin{align}
				\partial_t G = i ( G H - H^\dagger G ).
			\end{align}
			Therefore, taking the $q$-derivative of $G $, together with the Hermiticity of $G $ (i.e., $G = G^\dagger $), gives
			\begin{align}
				\partial_q G = X + X^\dagger ,
			\end{align}
			where $X $ is an operator to be determined. Since $G $ is a positive-definite operator by construction (hence invertible), we can always let $X = i G \widetilde{K} $ (we show its usefulness below) so that the $q$-derivative of $G $ becomes
			\begin{align}
				\partial_q G = i G \widetilde{K} - i \widetilde{K}^\dagger G .
			\end{align}
			Assuming physical states remain physical under a change of $q$, then $\Braket{\psi(p(s))}{\psi(p(s))} = 1$ for any $s$ with a continuous function ${p: s \mapsto (t, q(s))}$. Hence, the $s$-derivative of $\Braket{\psi(p(s))}{\psi(p(s))}$ should be zero, i.e.,

			\begin{align}
				0 = \frac{d}{ds} \Braket{\psi}{\psi} = \frac{dq}{ds} \partial_q \Braket{\psi}{\psi}.
			\end{align}
			For a path that satisfies $\dfrac{dq}{ds} \neq 0$ for all $s$, we obtain

			\begin{align}
				0 & = \partial_q \Braket{\psi }{\psi } = \partial_q \bra{\psi } G \ket{\psi }\\
				\Rightarrow & ~ \partial_q \ket{\psi } = - i \widetilde{K} \ket{\psi } + \ket{\zeta },
			\end{align}
			where $\ket{\zeta }$ is a state satisfying
			\begin{align}
				0 = \Braket{\zeta }{\psi } + \Braket{\psi }{\zeta }. \label{Real}
			\end{align}

					We can, therefore, decompose $\ket{\zeta }$ into two parts, i.e.,
			\begin{align}
				\ket{\zeta } = \ket{\xi } - i \widetilde{A} \ket{\psi }, \label{Zeta}
			\end{align}
			where $\ket{\xi }$ is independent of the input state $\ket{\psi }$, $\widetilde{A} $ is an operator, and $- i$ is for later convenience. Substituting Eq.~\eqref{Real} with Eq.~\eqref{Zeta} we find
			\begin{align}
				\begin{split}
					0 & = \left( \bra{\xi } + i \bra{\psi } \widetilde{A}^\dagger \right) G \ket{\psi } + \bra{\psi } G \left(\ket{\xi } - i \widetilde{A} \ket{\psi }\right)\\
					& = 2 \Re \bra{\xi } G \ket{\psi } + i \bra{\psi } \left( \widetilde{A}^\dagger G - G \widetilde{A} \right) \ket{\psi }.
				\end{split} \label{Xi1}
			\end{align}
			Since $\ket{\psi }$ is an arbitrary state, we can replace $\ket{\psi }$ with $\ket{\psi' } = \exp (i \theta) \ket{\psi }$, where $\theta \in (0, 2 \pi)$ which leads to
			\begin{align}
				\begin{split}
					0 & = 2 \Re \bra{\xi } G \ket{\psi' } + i \bra{\psi' } \left( \widetilde{A}^\dagger G - G \widetilde{A} \right) \ket{\psi' }\\
					& = 2 \Re \left( e^{i \theta} \bra{\xi } G \ket{\psi } \right) + i \bra{\psi } \left( \widetilde{A}^\dagger G - G \widetilde{A} \right) \ket{\psi }.
				\end{split} \label{Xi2}
			\end{align}
			Since $\theta$ is arbitrary, comparing Eqs.~\eqref{Xi1} and \eqref{Xi2}, we conclude that
			\begin{align}
				\bra{\xi } G \ket{\psi } = 0
			\end{align}
			for any $\ket{\psi }$. Moreover, since $G $ is positive definite, the only option left for $\ket{\zeta }$ is $\ket{\xi } = 0$ and $\widetilde{A}^\dagger G = G \widetilde{A} $.

					Therefore, the $q$-derivative of state $\ket{\psi }$ is
			\begin{align}
				\partial_q \ket{\psi } = - i \widetilde{K} \ket{\psi } - i \widetilde{A} \ket{\psi },
			\end{align}
			where $G \widetilde{A} = \widetilde{A}^\dagger G $. Combine $\tilde{K} $ and $\tilde{A} $ into a single operator $K = \tilde{K} + \tilde{A} $, so that
			\begin{align}
				\partial_q \ket{\psi } = - i K \ket{\psi },
			\end{align}
			and
			\begin{align}
				\begin{split}
					\partial_q G & = i G \widetilde{K} - i \widetilde{K}^\dagger G \\
					& = i G \widetilde{K} - i \widetilde{K}^\dagger G + i G \widetilde{A} - i G \widetilde{A} \\
					& = i G \widetilde{K} - i \widetilde{K}^\dagger G + i G \widetilde{A} - i \widetilde{A}^\dagger G \\
					& = i G K - i K^\dagger G .
				\end{split}
			\end{align}

			Hence, $K $ becomes the generator of both state and Hilbert space metric in the $q$-dimension. The constraints on $K $ can be found after the Hilbert space curvature is determined.

		\section{The adiabatic gauge\label{Appendix:AdiabaticGauge}}

			In this note, we show that the adiabatic gauge automatically leads to Eq.~\eqref{AdiabaticGaugeCondition}.

			Since $H (q)$ is time-independent, the eigenvalues remain the same while the eigenstates are evolving in time. Therefore, for an adiabatic process of any eigenstate, we have
			\begin{align}
				& H \ket{\psi_i } = h_i \ket{\psi_i }\\
				\Rightarrow & \quad \partial_q \big( H \ket{\psi_i } \big) = \partial_q \big( h_i \ket{\psi_i } \big) \nonumber\\
				\Rightarrow & \quad \big( \partial_q H - i H K {\big)} \ket{\psi_i (t,q)} = \big( \partial_q h_i - i h_i K \big) \ket{\psi_i } \nonumber\\
				\Rightarrow & \quad \big( \partial_q H - i H K {\big)} \ket{\psi_i } = \big( \partial_q h_i - i K H \big) \ket{\psi_i } \nonumber\\
				\Rightarrow & \quad \big( \partial_q H + i \left[ K , H \right] \big) \ket{\psi_i} = \big( \partial_q h_i \big) \ket{\psi_i } \nonumber\\
				\Rightarrow & \quad \big( \partial_t K \big) \ket{\psi_i } = \big( \partial_q h_i \big) \ket{\psi_i } \label{K_tEigenvalue}\\
				\Rightarrow & \quad \left[ \partial_t K , H \right] = 0. \label{AppendixMasterEquation2}
			\end{align}
			Equation~\eqref{AppendixMasterEquation2} results from the fact that $\partial_t K $ and $H $ share the same set of eigenstates. Note that Eq.~\eqref{K_tEigenvalue} shows that $\partial_t K $ acting on the eigenstates of $H $ gives the $q$-derivative of the corresponding eigenvalue.

		\section{Algebraic equations for $K$ in the adiabatic gauge\label{Appendix:AlgebraicEquations}}

			Here we show that for a time-independent Hamiltonian [$H = H (q)$], in the adiabatic gauge Eq.~\eqref{KEquation}, which is a differential equation, can be changed into algebraic equations. Algebraizations not only simplify the calculations, but very often, they make the underlying concept clearer~\cite{Kofman2023}. We start with Eq.~\eqref{KEquation} and the adiabatic gauge-fixing condition in Eq.~\eqref{AdiabaticGaugeCondition}
			\begin{align}
				& \partial_t K = i \left[ K , H \right] + \partial_q H , \label{IntegrableCond}\\
				& \left[ \partial_t K , H \right] = 0. \label{AGCond}
			\end{align}
			The first simplification comes from taking the time derivative of Eq.~\eqref{IntegrableCond}, we arrive at
			\begin{align}
				\partial_t^2 K & = i \left[ \partial_t K , H \right] + i \left[ K , \partial_t H \right] + \partial_t \partial_q H = i \left[ \partial_t K , H \right] = 0, \label{Alg}
			\end{align}
			where Eq.~\eqref{AGCond} have been applied in the last equality.

					Hence, Eqs.~\eqref{IntegrableCond}--\eqref{Alg} imply
			\begin{align}
				& K = t K^{(1)} + K^{(0)} , \label{K=At+B}\\
				& [K^{(1)} ,H ] = 0, \label{AZeroModes}\\
				& K^{(1)} = i \left[ K^{(0)} , H \right] + \partial_q H , \label{ABEq}
			\end{align}
			where $K^{(1)} $ and $K^{(0)} $ are time-independent matrices. Therefore, the differential equations in Eqs.~\eqref{IntegrableCond} and \eqref{AGCond} have now become the algebraic equations in Eqs.~\eqref{K=At+B}{--}\eqref{ABEq}.

			Further substituting Eq.~\eqref{K=At+B} into Eq.~\eqref{AGCond} gives Eq.~\eqref{AZeroModes}. That is to say, $K^{(1)} $ is composed of the zero modes of the operator $[ \boldsymbol{\cdot} , H ]$.

			Note that the residual gauge freedoms are always time independent. To verify this claim, we decompose $K^{(0)}$ into the zero modes and other modes, i.e.,
			\begin{align}
				K^{(0)} = K^{(0)}_\text{z} + K^{(0)}_\text{r} ,
			\end{align}
			where $K^{(0)}_\text{z} $ is the collection of the zero modes and $K^{(0)}_\text{r} $ is the rest. Then Eq.~\eqref{ABEq} becomes
			\begin{align}
				K^{(1)} & = i \left[ K^{(0)}_\text{z} + K^{(0)}_\text{r} , H \right] + \partial_q H = i \left[ K^{(0)}_\text{r} , H \right] + \partial_q H .
			\end{align}
			Since $K^{(1)} $ is composed of the zero modes and $[K^{(0)} , H ]$ is composed of the rest of the modes, where these two sets of modes are linearly independent, $\partial_q H $ can be fully and uniquely expressed by 
			\begin{align}
				\partial_q H = K^{(1)} - i [K^{(0)}_\text{r} , H ],
			\end{align}
			where the undetermined choices of $K^{(0)}_\text{z} $ are the gauge freedoms.

		\section{An example of the Berry curvature\label{Appendix:BerryCuravture}}

			To show the procedure of acquiring the Berry curvature from the generator $K$s, we demonstrate an example when a charged spin-1/2 particle is placed in a magnetic field with a constant magnitude, where the Hamiltonian of the particle takes the following form:
			\begin{align}
				H(\theta, \phi) = \mu \vec{B} \cdot \vec{\sigma} = \mu B \begin{pmatrix}
					\cos \theta & e^{-i \phi} \sin \theta\\
					e^{i \phi} \sin \theta & - \cos \theta
				\end{pmatrix},
			\end{align}
			where $\mu$ is the magnetic moment of the particle, $\vec{B}$ is the magnetic field, $\vec{\sigma}$ is the Pauli matrix vector, while $\theta$ and $\phi$ represent the direction of the magnetic field.

			Therefore, there are two parameters in the Hamiltonian, $\theta$ and $\phi$, which lead to the emergence of two additional dimensions in the evolution space. Using the adiabatic gauge, we find that the generators in the $\theta$ and $\phi$ directions are
			\begin{align}
				K_\theta & = \begin{pmatrix}
					\alpha_1 + \alpha_2 \cos \theta & \quad & e^{- i \phi} \bigg(- \dfrac{i}{2} + \alpha_2 \sin \theta \bigg)\\
					e^{i \phi} \bigg( \dfrac{i}{2} + \alpha_2 \sin \theta \bigg) & \quad & \alpha_1 - \alpha_2 \cos \theta
				\end{pmatrix}, \label{Ktheta}\\
				K_\phi & = \begin{pmatrix}
					\beta_1 + \beta_2 \cos \theta & \quad & e^{- i \phi} \bigg(- \dfrac{1}{2} \tan \theta + \beta_2 e^{i \phi} \sin \theta \bigg)\\
					e^{i \phi} \bigg( - \dfrac{1}{2} \tan \theta + \beta_2 e^{-i \phi} \sin \theta \bigg) & \quad & \beta_1 - \beta_2 \cos \theta
				\end{pmatrix}, \label{Kphi}
			\end{align}
			where the generators are time-independent and the undetermined functions of $(\theta, \phi)$, $\alpha_{1 / 2} $ and $\beta_{1 / 2} $, also correspond to gauge freedom. Although it seems that there are four gauge degrees of freedom, these functions are related through Eq.~\eqref{MultiKEquation2}, namely,
			\begin{align}
				& \partial_\phi \alpha_1 = \partial_\theta \beta_1 , \label{GaugeEq1}\\
				& 2 \cos^2 \theta \big( \partial_\theta \beta_2 - \partial_\phi \alpha_2 \big) = \sin \theta, \label{GaugeEq2}
			\end{align}
			so that there are only two degrees of freedom left.

			Nevertheless, as stated in the main article, the residual gauge transformation does not affect the Berry curvature. Therefore, we insert Eqs.~\eqref{Ktheta} and \eqref{Kphi} without the need of solving Eqs.~\eqref{GaugeEq1} and \eqref{GaugeEq2} into
			\begin{align}
				\Omega_{\theta \phi}^\pm = i \bra{\psi_\pm} \left[ K_\theta , K_\phi \right] \ket{\psi_\pm} = \mp \frac{\sin \theta}{2}, \label{BerryCurvature}
			\end{align}
			where
			\begin{align}
				& \ket{\psi_-} = \begin{pmatrix} \sin \left( \theta / 2 \right) \exp( -i \phi)\\ - \cos \left( \theta / 2 \right) \end{pmatrix},\\
				& \ket{\psi_+} = \begin{pmatrix} \cos \left( \theta / 2 \right) \exp( -i \phi)\\ \sin \left( \theta / 2 \right) \end{pmatrix},
			\end{align}
			are the eigenstates of the Hamiltonian $H$.

			Note that the result in Eq.~\eqref{BerryCurvature} is indeed independent of $\alpha_{1 / 2} $ and $\beta_{1 / 2} $ (i.e., the gauge choices) and is consistent with the well-known example in Ref.~\cite{Xiao2010}.

		\section{An example of fidelity susceptibility\label{Appendix:FidelitySusceptibility}}

			In this note we demonstrate how the fidelity susceptibility works using the example provided in the main article, namely,
			\begin{align}
				H( \gamma ) = \begin{pmatrix}
					i \gamma & 1\\
					1 & -i \gamma
				\end{pmatrix},
			\end{align}
			with $\gamma$ being the parameter to be investigated. Note that the Hamiltonian is at an exceptional point when $\gamma = \pm 1$.

					The $\gamma$-direction generator $K $ has been found to be
			\begin{align}
				K & = \dfrac{1}{2\left(\gamma^2 - 1\right)}\begin{pmatrix}
					2 i \gamma^2 t & 2 \gamma t + 1\\
					2 \gamma t - 1 & - 2 i \gamma^2 t
				\end{pmatrix} + C,
			\end{align}
			where
			\begin{align}
				C & = \begin{pmatrix}
					c_1 + i \gamma c_2 & c_2 \\
					c_2 & c_1 - i \gamma c_2 
				\end{pmatrix},
			\end{align}
			with $c_1 $ and $c_2 $ being the undetermined functions corresponding to two gauge degrees of freedom. The generalized fidelity susceptibility~\cite{Tzeng2021},
			\begin{align}
				\chi_n & = \Bra{\psi_n } K^2 \Ket{\psi_n } - \Bra{\psi_n } K \Ket{\psi_n }^2,
			\end{align}
			can be found, where $\Ket{\psi_n } = \ket{\psi_n }$ is an eigenstate of the Hamiltonian, namely,
			\begin{align}
				\ket{\psi_\pm } = e^{\mp i \epsilon}\begin{pmatrix}
					(\epsilon \pm i \gamma )^{1/2}\\
					(\pm \epsilon - i \gamma )^{1/2}
				\end{pmatrix},
			\end{align}
			where $\epsilon = \sqrt{1 - \gamma^2}$.

			Combining all the information together, we find that the fidelity susceptibilities of both eigenstates, $\ket{\psi_\pm }$, are
			\begin{align}
				\chi_\pm = \frac{-1}{4(1 - \gamma^2)^2},
			\end{align}
			which are indeed independent of time and gauge choices, and are singular at the Hamiltonian exceptional points. Furthermore, this result is the same as the one found in the literature~\cite{Tzeng2021}.
		\end{appendix}

		\newpage

		\bibliography{References}

\begin{thebibliography}{69}%
\makeatletter
\providecommand \@ifxundefined [1]{%
 \@ifx{#1\undefined}
}%
\providecommand \@ifnum [1]{%
 \ifnum #1\expandafter \@firstoftwo
 \else \expandafter \@secondoftwo
 \fi
}%
\providecommand \@ifx [1]{%
 \ifx #1\expandafter \@firstoftwo
 \else \expandafter \@secondoftwo
 \fi
}%
\providecommand \natexlab [1]{#1}%
\providecommand \enquote  [1]{``#1''}%
\providecommand \bibnamefont  [1]{#1}%
\providecommand \bibfnamefont [1]{#1}%
\providecommand \citenamefont [1]{#1}%
\providecommand \href@noop [0]{\@secondoftwo}%
\providecommand \href [0]{\begingroup \@sanitize@url \@href}%
\providecommand \@href[1]{\@@startlink{#1}\@@href}%
\providecommand \@@href[1]{\endgroup#1\@@endlink}%
\providecommand \@sanitize@url [0]{\catcode `\\12\catcode `\$12\catcode
  `\&12\catcode `\#12\catcode `\^12\catcode `\_12\catcode `\%12\relax}%
\providecommand \@@startlink[1]{}%
\providecommand \@@endlink[0]{}%
\providecommand \url  [0]{\begingroup\@sanitize@url \@url }%
\providecommand \@url [1]{\endgroup\@href {#1}{\urlprefix }}%
\providecommand \urlprefix  [0]{URL }%
\providecommand \Eprint [0]{\href }%
\providecommand \doibase [0]{http://dx.doi.org/}%
\providecommand \selectlanguage [0]{\@gobble}%
\providecommand \bibinfo  [0]{\@secondoftwo}%
\providecommand \bibfield  [0]{\@secondoftwo}%
\providecommand \translation [1]{[#1]}%
\providecommand \BibitemOpen [0]{}%
\providecommand \bibitemStop [0]{}%
\providecommand \bibitemNoStop [0]{.\EOS\space}%
\providecommand \EOS [0]{\spacefactor3000\relax}%
\providecommand \BibitemShut  [1]{\csname bibitem#1\endcsname}%
\let\auto@bib@innerbib\@empty
\bibitem [{\citenamefont {Bender}\ and\ \citenamefont
  {Boettcher}(1998)}]{Bender1998}%
  \BibitemOpen
  \bibinfo {author} {C.~M. Bender}\ and\ \bibinfo {author} {S.~Boettcher},\
  \emph {\bibinfo {title} {Real Spectra in Non-{H}ermitian {H}amiltonians
  Having $\mathcal{PT}$ Symmetry}},\ \href {\doibase
  10.1103/PhysRevLett.80.5243} {\bibfield  {journal} {\bibinfo  {journal}
  {Phys. Rev. Lett.}\ }\textbf {\bibinfo {volume} {80}},\ \bibinfo {pages}
  {5243} (\bibinfo {year} {1998})}\BibitemShut {NoStop}%
\bibitem [{\citenamefont {Bender}(2007)}]{Bender2007}%
  \BibitemOpen
  \bibinfo {author} {C.~M. Bender},\ \emph {\bibinfo {title} {Making sense of
  non-{H}ermitian {H}amiltonians}},\ \href {\doibase
  10.1088/0034-4885/70/6/R03} {\bibfield  {journal} {\bibinfo  {journal} {Rep.
  Prog. Phys.}\ }\textbf {\bibinfo {volume} {70}},\ \bibinfo {pages} {947}
  (\bibinfo {year} {2007})}\BibitemShut {NoStop}%
\bibitem [{\citenamefont {Makris}\ \emph {et~al.}(2008)\citenamefont {Makris},
  \citenamefont {El-Ganainy}, \citenamefont {Christodoulides},\ and\
  \citenamefont {Musslimani}}]{Makris2008}%
  \BibitemOpen
  \bibinfo {author} {K.~G. Makris}, \bibinfo {author} {R.~El-Ganainy}, \bibinfo
  {author} {D.~N. Christodoulides},\ and\ \bibinfo {author} {Z.~H.
  Musslimani},\ \emph {\bibinfo {title} {Beam Dynamics in $\cal{PT}$ Symmetric
  Optical Lattices}},\ \href {\doibase 10.1103/physrevlett.100.103904}
  {\bibfield  {journal} {\bibinfo  {journal} {Phys. Rev. Lett.}\ }\textbf
  {\bibinfo {volume} {100}},\ \bibinfo {pages} {103904} (\bibinfo {year}
  {2008})}\BibitemShut {NoStop}%
\bibitem [{\citenamefont {El-Ganainy}\ \emph {et~al.}(2018)\citenamefont
  {El-Ganainy}, \citenamefont {Makris}, \citenamefont {Khajavikhan},
  \citenamefont {Musslimani}, \citenamefont {Rotter},\ and\ \citenamefont
  {Christodoulides}}]{ElGanainy2018}%
  \BibitemOpen
  \bibinfo {author} {R.~El-Ganainy}, \bibinfo {author} {K.~G. Makris}, \bibinfo
  {author} {M.~Khajavikhan}, \bibinfo {author} {Z.~H. Musslimani}, \bibinfo
  {author} {S.~Rotter},\ and\ \bibinfo {author} {D.~N. Christodoulides},\ \emph
  {\bibinfo {title} {Non-Hermitian physics and {$\cal{PT}$} symmetry}},\ \href
  {\doibase 10.1038/nphys4323} {\bibfield  {journal} {\bibinfo  {journal} {Nat.
  Phys.}\ }\textbf {\bibinfo {volume} {14}},\ \bibinfo {pages} {11} (\bibinfo
  {year} {2018})}\BibitemShut {NoStop}%
\bibitem [{\citenamefont {Mostafazadeh}(2003)}]{Mostafazadeh2003}%
  \BibitemOpen
  \bibinfo {author} {A.~Mostafazadeh},\ \emph {\bibinfo {title}
  {Pseudo-{H}ermiticity and generalized $\mathcal{PT}$- and
  $\mathcal{CPT}$-symmetries}},\ \href {\doibase 10.1063/1.1539304} {\bibfield
  {journal} {\bibinfo  {journal} {J. Math. Phys.}\ }\textbf {\bibinfo {volume}
  {44}},\ \bibinfo {pages} {974} (\bibinfo {year} {2003})}\BibitemShut
  {NoStop}%
\bibitem [{\citenamefont {Mostafazadeh}(2010)}]{Mostafazadeh2010}%
  \BibitemOpen
  \bibinfo {author} {A.~Mostafazadeh},\ \emph {\bibinfo {title}
  {Pseudo-Hermitian representation of quantum mechanics}},\ \href {\doibase
  10.1142/S0219887810004816} {\bibfield  {journal} {\bibinfo  {journal} {Int.
  J. Geom. Meth. Mod. Phys.}\ }\textbf {\bibinfo {volume} {7}},\ \bibinfo
  {pages} {1191} (\bibinfo {year} {2010})}\BibitemShut {NoStop}%
\bibitem [{\citenamefont {Peng}\ \emph {et~al.}(2014)\citenamefont {Peng},
  \citenamefont {Özdemir}, \citenamefont {Rotter}, \citenamefont {Yilmaz},
  \citenamefont {Liertzer}, \citenamefont {Monifi}, \citenamefont {Bender},
  \citenamefont {Nori},\ and\ \citenamefont {Yang}}]{Peng2014}%
  \BibitemOpen
  \bibinfo {author} {B.~Peng}, \bibinfo {author} {{\c{S}}.~K. Özdemir},
  \bibinfo {author} {S.~Rotter}, \bibinfo {author} {H.~Yilmaz}, \bibinfo
  {author} {M.~Liertzer}, \bibinfo {author} {F.~Monifi}, \bibinfo {author}
  {C.~M. Bender}, \bibinfo {author} {F.~Nori},\ and\ \bibinfo {author}
  {L.~Yang},\ \emph {\bibinfo {title} {Loss-induced suppression and revival of
  lasing}},\ \href {\doibase 10.1126/science.1258004} {\bibfield  {journal}
  {\bibinfo  {journal} {Science}\ }\textbf {\bibinfo {volume} {346}},\ \bibinfo
  {pages} {328} (\bibinfo {year} {2014})}\BibitemShut {NoStop}%
\bibitem [{\citenamefont {Jing}\ \emph {et~al.}(2014)\citenamefont {Jing},
  \citenamefont {Özdemir}, \citenamefont {Lü}, \citenamefont {Zhang},
  \citenamefont {Yang},\ and\ \citenamefont {Nori}}]{Jing2014}%
  \BibitemOpen
  \bibinfo {author} {H.~Jing}, \bibinfo {author} {{\c{S}}.~K. Özdemir},
  \bibinfo {author} {X.-Y. Lü}, \bibinfo {author} {J.~Zhang}, \bibinfo
  {author} {L.~Yang},\ and\ \bibinfo {author} {F.~Nori},\ \emph {\bibinfo
  {title} {{$\cal{PT}$}-Symmetric Phonon Laser}},\ \href {\doibase
  10.1103/physrevlett.113.053604} {\bibfield  {journal} {\bibinfo  {journal}
  {Phys. Rev. Lett.}\ }\textbf {\bibinfo {volume} {113}},\ \bibinfo {pages}
  {053604} (\bibinfo {year} {2014})}\BibitemShut {NoStop}%
\bibitem [{\citenamefont {Bender}(2016)}]{Bender2016}%
  \BibitemOpen
  \bibinfo {author} {C.~M. Bender},\ \emph {\bibinfo {title} {{$\cal{PT}$}
  symmetry in quantum physics: From a mathematical curiosity to optical
  experiments}},\ \href {\doibase 10.1051/epn/2016201} {\bibfield  {journal}
  {\bibinfo  {journal} {Europhys. News}\ }\textbf {\bibinfo {volume} {47}},\
  \bibinfo {pages} {17} (\bibinfo {year} {2016})}\BibitemShut {NoStop}%
\bibitem [{\citenamefont {Bender}\ \emph {et~al.}(2017)\citenamefont {Bender},
  \citenamefont {Brody},\ and\ \citenamefont {Müller}}]{Bender2017}%
  \BibitemOpen
  \bibinfo {author} {C.~M. Bender}, \bibinfo {author} {D.~C. Brody},\ and\
  \bibinfo {author} {M.~P. Müller},\ \emph {\bibinfo {title} {Hamiltonian for
  the Zeros of the Riemann Zeta Function}},\ \href {\doibase
  10.1103/physrevlett.118.130201} {\bibfield  {journal} {\bibinfo  {journal}
  {Phys. Rev. Lett.}\ }\textbf {\bibinfo {volume} {118}},\ \bibinfo {pages}
  {130201} (\bibinfo {year} {2017})}\BibitemShut {NoStop}%
\bibitem [{\citenamefont {Miller}(2017)}]{Miller2017}%
  \BibitemOpen
  \bibinfo {author} {J.~L. Miller},\ \emph {\bibinfo {title} {Exceptional
  points make for exceptional sensors}},\ \href {\doibase 10.1063/pt.3.3717}
  {\bibfield  {journal} {\bibinfo  {journal} {Phys. Today}\ }\textbf {\bibinfo
  {volume} {70}},\ \bibinfo {pages} {23} (\bibinfo {year} {2017})}\BibitemShut
  {NoStop}%
\bibitem [{\citenamefont {Leykam}\ \emph {et~al.}(2017)\citenamefont {Leykam},
  \citenamefont {Bliokh}, \citenamefont {Huang}, \citenamefont {Chong},\ and\
  \citenamefont {Nori}}]{Leykam2017}%
  \BibitemOpen
  \bibinfo {author} {D.~Leykam}, \bibinfo {author} {K.~Y. Bliokh}, \bibinfo
  {author} {C.~Huang}, \bibinfo {author} {Y.~Chong},\ and\ \bibinfo {author}
  {F.~Nori},\ \emph {\bibinfo {title} {Edge Modes, Degeneracies, and
  Topological Numbers in Non-Hermitian Systems}},\ \href {\doibase
  10.1103/physrevlett.118.040401} {\bibfield  {journal} {\bibinfo  {journal}
  {Phys. Rev. Lett.}\ }\textbf {\bibinfo {volume} {118}},\ \bibinfo {pages}
  {040401} (\bibinfo {year} {2017})}\BibitemShut {NoStop}%
\bibitem [{\citenamefont {Quijandr{\'{\i}}a}\ \emph {et~al.}(2018)\citenamefont
  {Quijandr{\'{\i}}a}, \citenamefont {Naether}, \citenamefont {Özdemir},
  \citenamefont {Nori},\ and\ \citenamefont {Zueco}}]{Quijandria2018}%
  \BibitemOpen
  \bibinfo {author} {F.~Quijandr{\'{\i}}a}, \bibinfo {author} {U.~Naether},
  \bibinfo {author} {S.~K. Özdemir}, \bibinfo {author} {F.~Nori},\ and\
  \bibinfo {author} {D.~Zueco},\ \emph {\bibinfo {title}
  {{$\cal{PT}$}-symmetric circuit {QED}}},\ \href {\doibase
  10.1103/physreva.97.053846} {\bibfield  {journal} {\bibinfo  {journal} {Phys.
  Rev. A}\ }\textbf {\bibinfo {volume} {97}},\ \bibinfo {pages} {053846}
  (\bibinfo {year} {2018})}\BibitemShut {NoStop}%
\bibitem [{\citenamefont {El-Ganainy}\ \emph {et~al.}(2019)\citenamefont
  {El-Ganainy}, \citenamefont {Khajavikhan}, \citenamefont {Christodoulides},\
  and\ \citenamefont {Özdemir}}]{ElGanainy2019}%
  \BibitemOpen
  \bibinfo {author} {R.~El-Ganainy}, \bibinfo {author} {M.~Khajavikhan},
  \bibinfo {author} {D.~N. Christodoulides},\ and\ \bibinfo {author}
  {{\c{S}}.~K. Özdemir},\ \emph {\bibinfo {title} {The dawn of non-Hermitian
  optics}},\ \href {\doibase 10.1038/s42005-019-0130-z} {\bibfield  {journal}
  {\bibinfo  {journal} {Commun. Phys.}\ }\textbf {\bibinfo {volume} {2}},\
  \bibinfo {pages} {37} (\bibinfo {year} {2019})}\BibitemShut {NoStop}%
\bibitem [{\citenamefont {Liu}\ \emph {et~al.}(2019)\citenamefont {Liu},
  \citenamefont {Zhang}, \citenamefont {Ai}, \citenamefont {Gong},
  \citenamefont {Kawabata}, \citenamefont {Ueda},\ and\ \citenamefont
  {Nori}}]{Liu2019}%
  \BibitemOpen
  \bibinfo {author} {T.~Liu}, \bibinfo {author} {Y.-R. Zhang}, \bibinfo
  {author} {Q.~Ai}, \bibinfo {author} {Z.~Gong}, \bibinfo {author}
  {K.~Kawabata}, \bibinfo {author} {M.~Ueda},\ and\ \bibinfo {author}
  {F.~Nori},\ \emph {\bibinfo {title} {Second-Order Topological Phases in
  Non-Hermitian Systems}},\ \href {\doibase 10.1103/physrevlett.122.076801}
  {\bibfield  {journal} {\bibinfo  {journal} {Phys. Rev. Lett.}\ }\textbf
  {\bibinfo {volume} {122}},\ \bibinfo {pages} {076801} (\bibinfo {year}
  {2019})}\BibitemShut {NoStop}%
\bibitem [{\citenamefont {Ge}\ \emph {et~al.}(2019)\citenamefont {Ge},
  \citenamefont {Zhang}, \citenamefont {Liu}, \citenamefont {Li}, \citenamefont
  {Fan},\ and\ \citenamefont {Nori}}]{Ge2019}%
  \BibitemOpen
  \bibinfo {author} {Z.-Y. Ge}, \bibinfo {author} {Y.-R. Zhang}, \bibinfo
  {author} {T.~Liu}, \bibinfo {author} {S.-W. Li}, \bibinfo {author} {H.~Fan},\
  and\ \bibinfo {author} {F.~Nori},\ \emph {\bibinfo {title} {Topological band
  theory for non-Hermitian systems from the Dirac equation}},\ \href {\doibase
  10.1103/physrevb.100.054105} {\bibfield  {journal} {\bibinfo  {journal}
  {Phys. Rev. B}\ }\textbf {\bibinfo {volume} {100}},\ \bibinfo {pages}
  {054105} (\bibinfo {year} {2019})}\BibitemShut {NoStop}%
\bibitem [{\citenamefont {Parto}\ \emph {et~al.}(2020)\citenamefont {Parto},
  \citenamefont {Liu}, \citenamefont {Bahari}, \citenamefont {Khajavikhan},\
  and\ \citenamefont {Christodoulides}}]{Parto2020}%
  \BibitemOpen
  \bibinfo {author} {M.~Parto}, \bibinfo {author} {Y.~G.~N. Liu}, \bibinfo
  {author} {B.~Bahari}, \bibinfo {author} {M.~Khajavikhan},\ and\ \bibinfo
  {author} {D.~N. Christodoulides},\ \emph {\bibinfo {title} {Non-Hermitian and
  topological photonics: optics at an exceptional point}},\ \href {\doibase
  10.1515/nanoph-2020-0434} {\bibfield  {journal} {\bibinfo  {journal} {P. Soc.
  Photo-opt. Ins.}\ }\textbf {\bibinfo {volume} {10}},\ \bibinfo {pages} {403}
  (\bibinfo {year} {2020})}\BibitemShut {NoStop}%
\bibitem [{\citenamefont {Ashida}\ \emph {et~al.}(2020)\citenamefont {Ashida},
  \citenamefont {Gong},\ and\ \citenamefont {Ueda}}]{Ashida2020}%
  \BibitemOpen
  \bibinfo {author} {Y.~Ashida}, \bibinfo {author} {Z.~Gong},\ and\ \bibinfo
  {author} {M.~Ueda},\ \emph {\bibinfo {title} {Non-Hermitian physics}},\ \href
  {\doibase 10.1080/00018732.2021.1876991} {\bibfield  {journal} {\bibinfo
  {journal} {Adv. Phys.}\ }\textbf {\bibinfo {volume} {69}},\ \bibinfo {pages}
  {249} (\bibinfo {year} {2020})}\BibitemShut {NoStop}%
\bibitem [{\citenamefont {Cirio}\ \emph {et~al.}(2022)\citenamefont {Cirio},
  \citenamefont {Kuo}, \citenamefont {Chen}, \citenamefont {Nori},\ and\
  \citenamefont {Lambert}}]{Cirio2022}%
  \BibitemOpen
  \bibinfo {author} {M.~Cirio}, \bibinfo {author} {P.-C. Kuo}, \bibinfo
  {author} {Y.-N. Chen}, \bibinfo {author} {F.~Nori},\ and\ \bibinfo {author}
  {N.~Lambert},\ \emph {\bibinfo {title} {Canonical derivation of the fermionic
  influence superoperator}},\ \href {\doibase 10.1103/physrevb.105.035121}
  {\bibfield  {journal} {\bibinfo  {journal} {Phys. Rev. B}\ }\textbf {\bibinfo
  {volume} {105}},\ \bibinfo {pages} {035121} (\bibinfo {year}
  {2022})}\BibitemShut {NoStop}%
\bibitem [{\citenamefont {Bergholtz}\ \emph {et~al.}(2021)\citenamefont
  {Bergholtz}, \citenamefont {Budich},\ and\ \citenamefont
  {Kunst}}]{Bergholtz2021}%
  \BibitemOpen
  \bibinfo {author} {E.~J. Bergholtz}, \bibinfo {author} {J.~C. Budich},\ and\
  \bibinfo {author} {F.~K. Kunst},\ \emph {\bibinfo {title} {Exceptional
  topology of non-Hermitian systems}},\ \href {\doibase
  10.1103/revmodphys.93.015005} {\bibfield  {journal} {\bibinfo  {journal}
  {Rev. Mod. Phys.}\ }\textbf {\bibinfo {volume} {93}},\ \bibinfo {pages}
  {015005} (\bibinfo {year} {2021})}\BibitemShut {NoStop}%
\bibitem [{\citenamefont {Zhang}\ \emph {et~al.}(2022)\citenamefont {Zhang},
  \citenamefont {Zhang}, \citenamefont {Lu},\ and\ \citenamefont
  {Chen}}]{Zhang2022}%
  \BibitemOpen
  \bibinfo {author} {X.~Zhang}, \bibinfo {author} {T.~Zhang}, \bibinfo {author}
  {M.-H. Lu},\ and\ \bibinfo {author} {Y.-F. Chen},\ \emph {\bibinfo {title} {A
  review on non-Hermitian skin effect}},\ \href {\doibase
  10.1080/23746149.2022.2109431} {\bibfield  {journal} {\bibinfo  {journal}
  {Adva. Phys.: X}\ }\textbf {\bibinfo {volume} {7}},\ \bibinfo {pages}
  {2109431} (\bibinfo {year} {2022})}\BibitemShut {NoStop}%
\bibitem [{\citenamefont {Fring}(2023)}]{Fring2022}%
  \BibitemOpen
  \bibinfo {author} {A.~Fring},\ \emph {\bibinfo {title} {An Introduction to
  PT-Symmetric Quantum Mechanics-Time-Dependent Systems}},\ \href {\doibase
  10.1088/1742-6596/2448/1/012002} {\bibfield  {journal} {\bibinfo  {journal}
  {J. Phys.: Conf. Ser.}\ }\textbf {\bibinfo {volume} {2448}},\ \bibinfo
  {pages} {012002} (\bibinfo {year} {2023})}\BibitemShut {NoStop}%
\bibitem [{\citenamefont {Fang}\ \emph {et~al.}(2022)\citenamefont {Fang},
  \citenamefont {Zhao}, \citenamefont {Chen}, \citenamefont {Zhou},
  \citenamefont {Zhang}, \citenamefont {Wu}, \citenamefont {Yang},\ and\
  \citenamefont {Nori}}]{Fang2022}%
  \BibitemOpen
  \bibinfo {author} {Y.-L. Fang}, \bibinfo {author} {J.-L. Zhao}, \bibinfo
  {author} {D.-X. Chen}, \bibinfo {author} {Y.-H. Zhou}, \bibinfo {author}
  {Y.~Zhang}, \bibinfo {author} {Q.-C. Wu}, \bibinfo {author} {C.-P. Yang},\
  and\ \bibinfo {author} {F.~Nori},\ \emph {\bibinfo {title} {Entanglement
  dynamics in anti-{$\cal{PT}$}-symmetric systems}},\ \href {\doibase
  10.1103/physrevresearch.4.033022} {\bibfield  {journal} {\bibinfo  {journal}
  {Phys. Rev. Research}\ }\textbf {\bibinfo {volume} {4}},\ \bibinfo {pages}
  {033022} (\bibinfo {year} {2022})}\BibitemShut {NoStop}%
\bibitem [{\citenamefont {Chen}\ \emph {et~al.}(2022)\citenamefont {Chen},
  \citenamefont {Zhang}, \citenamefont {Zhao}, \citenamefont {Wu},
  \citenamefont {Fang}, \citenamefont {Yang},\ and\ \citenamefont
  {Nori}}]{Chen2022}%
  \BibitemOpen
  \bibinfo {author} {D.-X. Chen}, \bibinfo {author} {Y.~Zhang}, \bibinfo
  {author} {J.-L. Zhao}, \bibinfo {author} {Q.-C. Wu}, \bibinfo {author} {Y.-L.
  Fang}, \bibinfo {author} {C.-P. Yang},\ and\ \bibinfo {author} {F.~Nori},\
  \emph {\bibinfo {title} {Quantum state discrimination in a
  {$\cal{PT}$}-symmetric system}},\ \href {\doibase
  10.1103/physreva.106.022438} {\bibfield  {journal} {\bibinfo  {journal}
  {Phys. Rev. A}\ }\textbf {\bibinfo {volume} {106}},\ \bibinfo {pages}
  {022438} (\bibinfo {year} {2022})}\BibitemShut {NoStop}%
\bibitem [{\citenamefont {Fring}\ and\ \citenamefont {Taira}()}]{Fring2023}%
  \BibitemOpen
  \bibinfo {author} {A.~Fring}\ and\ \bibinfo {author} {T.~Taira},\ \emph
  {\bibinfo {title} {Non-Hermitian quantum Fermi accelerator}},\ \href
  {\doibase 10.1103/physreva.108.012222} {\bibfield  {journal} {\bibinfo
  {journal} {Phys. Rev. A}\ }\textbf {\bibinfo {volume} {108}},\
  10.1103/physreva.108.012222}\BibitemShut {NoStop}%
\bibitem [{\citenamefont {Znojil}(2024)}]{Znojil2024}%
  \BibitemOpen
  \bibinfo {author} {M.~Znojil},\ \emph {\bibinfo {title} {Discrete-coordinate
  crypto-Hermitian quantum system controlled by time-dependent Robin boundary
  conditions}},\ \href {\doibase 10.1088/1402-4896/ad298b} {\bibfield
  {journal} {\bibinfo  {journal} {Phys. Scripta}\ }\textbf {\bibinfo {volume}
  {99}},\ \bibinfo {pages} {035250} (\bibinfo {year} {2024})}\BibitemShut
  {NoStop}%
\bibitem [{\citenamefont {Znojil}(2008)}]{Znojil2008}%
  \BibitemOpen
  \bibinfo {author} {M.~Znojil},\ \emph {\bibinfo {title} {Time-dependent
  version of crypto-Hermitian quantum theory}},\ \href {\doibase
  10.1103/PhysRevD.78.085003} {\bibfield  {journal} {\bibinfo  {journal} {Phys.
  Rev. D}\ }\textbf {\bibinfo {volume} {78}},\ \bibinfo {pages} {085003}
  (\bibinfo {year} {2008})}\BibitemShut {NoStop}%
\bibitem [{\citenamefont {Znojil}(2009)}]{Znojil2009}%
  \BibitemOpen
  \bibinfo {author} {M.~Znojil},\ \emph {\bibinfo {title} {Three-Hilbert-Space
  Formulation of Quantum Mechanics}},\ \href {\doibase 10.3842/sigma.2009.001}
  {\bibfield  {journal} {\bibinfo  {journal} {Sym. Integ. Geom.: Meth. App.}\
  }\textbf {\bibinfo {volume} {5}},\ \bibinfo {eid} {001} (\bibinfo {year}
  {2009})}\BibitemShut {NoStop}%
\bibitem [{\citenamefont {Brody}(2013)}]{Brody2013}%
  \BibitemOpen
  \bibinfo {author} {D.~C. Brody},\ \emph {\bibinfo {title} {Biorthogonal
  quantum mechanics}},\ \href {\doibase 10.1088/1751-8113/47/3/035305}
  {\bibfield  {journal} {\bibinfo  {journal} {J. Phys. A: Math. Theor.}\
  }\textbf {\bibinfo {volume} {47}},\ \bibinfo {pages} {035305} (\bibinfo
  {year} {2013})}\BibitemShut {NoStop}%
\bibitem [{\citenamefont {Hodaei}\ \emph {et~al.}(2017)\citenamefont {Hodaei},
  \citenamefont {Hassan}, \citenamefont {Wittek}, \citenamefont
  {Garcia-Gracia}, \citenamefont {El-Ganainy}, \citenamefont
  {Christodoulides},\ and\ \citenamefont {Khajavikhan}}]{Hodaei2017}%
  \BibitemOpen
  \bibinfo {author} {H.~Hodaei}, \bibinfo {author} {A.~U. Hassan}, \bibinfo
  {author} {S.~Wittek}, \bibinfo {author} {H.~Garcia-Gracia}, \bibinfo {author}
  {R.~El-Ganainy}, \bibinfo {author} {D.~N. Christodoulides},\ and\ \bibinfo
  {author} {M.~Khajavikhan},\ \emph {\bibinfo {title} {Enhanced sensitivity at
  higher-order exceptional points}},\ \href {\doibase 10.1038/nature23280}
  {\bibfield  {journal} {\bibinfo  {journal} {Nature (London)}\ }\textbf
  {\bibinfo {volume} {548}},\ \bibinfo {pages} {187} (\bibinfo {year}
  {2017})}\BibitemShut {NoStop}%
\bibitem [{\citenamefont {Bliokh}\ \emph {et~al.}(2019)\citenamefont {Bliokh},
  \citenamefont {Leykam}, \citenamefont {Lein},\ and\ \citenamefont
  {Nori}}]{Bliokh2019}%
  \BibitemOpen
  \bibinfo {author} {K.~Y. Bliokh}, \bibinfo {author} {D.~Leykam}, \bibinfo
  {author} {M.~Lein},\ and\ \bibinfo {author} {F.~Nori},\ \emph {\bibinfo
  {title} {Topological non-Hermitian origin of surface Maxwell waves}},\ \href
  {\doibase 10.1038/s41467-019-08397-6} {\bibfield  {journal} {\bibinfo
  {journal} {Nat. Commun.}\ }\textbf {\bibinfo {volume} {10}},\ \bibinfo
  {pages} {580} (\bibinfo {year} {2019})}\BibitemShut {NoStop}%
\bibitem [{\citenamefont {Znojil}(2020)}]{Znojil2020}%
  \BibitemOpen
  \bibinfo {author} {M.~Znojil},\ \emph {\bibinfo {title} {Passage through
  exceptional point: Case study}},\ \href {\doibase 10.1098/rspa.2019.0831}
  {\bibfield  {journal} {\bibinfo  {journal} {Proc. Royal Soc. A}\ }\textbf
  {\bibinfo {volume} {476}},\ \bibinfo {pages} {20190831} (\bibinfo {year}
  {2020})}\BibitemShut {NoStop}%
\bibitem [{\citenamefont {Znojil}(2021)}]{Znojil2021}%
  \BibitemOpen
  \bibinfo {author} {M.~Znojil},\ \emph {\bibinfo {title} {Paths of unitary
  access to exceptional points}},\ \href {\doibase
  10.1088/1742-6596/2038/1/012026} {\bibfield  {journal} {\bibinfo  {journal}
  {J. Phys.: Conf. Ser.}\ }\textbf {\bibinfo {volume} {2038}},\ \bibinfo
  {pages} {012026} (\bibinfo {year} {2021})}\BibitemShut {NoStop}%
\bibitem [{\citenamefont {Bender}\ \emph {et~al.}(2004)\citenamefont {Bender},
  \citenamefont {Brod}, \citenamefont {Refig},\ and\ \citenamefont
  {Reuter}}]{Bender2004}%
  \BibitemOpen
  \bibinfo {author} {C.~M. Bender}, \bibinfo {author} {J.~Brod}, \bibinfo
  {author} {A.~Refig},\ and\ \bibinfo {author} {M.~E. Reuter},\ \emph {\bibinfo
  {title} {The $\mathcal{C}$ operator in $\mathcal{PT}$-symmetric quantum
  theories}},\ \href {\doibase 10.1088/0305-4470/37/43/009} {\bibfield
  {journal} {\bibinfo  {journal} {J. Phys A: Math. Gen.}\ }\textbf {\bibinfo
  {volume} {37}},\ \bibinfo {pages} {10139} (\bibinfo {year}
  {2004})}\BibitemShut {NoStop}%
\bibitem [{\citenamefont {Mostafazadeh}(2004)}]{Mostafazadeh2004}%
  \BibitemOpen
  \bibinfo {author} {A.~Mostafazadeh},\ \emph {\bibinfo {title} {{Time
  dependent Hilbert spaces, geometric phases, and general covariance in quantum
  mechanics}}},\ \href {\doibase 10.1016/j.physleta.2003.12.008} {\bibfield
  {journal} {\bibinfo  {journal} {Phys. Lett. A}\ }\textbf {\bibinfo {volume}
  {320}},\ \bibinfo {pages} {375} (\bibinfo {year} {2004})}\BibitemShut
  {NoStop}%
\bibitem [{\citenamefont {Ju}\ \emph {et~al.}(2022)\citenamefont {Ju},
  \citenamefont {Miranowicz}, \citenamefont {Minganti}, \citenamefont {Chan},
  \citenamefont {Chen},\ and\ \citenamefont {Nori}}]{Ju2021}%
  \BibitemOpen
  \bibinfo {author} {C.-Y. Ju}, \bibinfo {author} {A.~Miranowicz}, \bibinfo
  {author} {F.~Minganti}, \bibinfo {author} {C.-T. Chan}, \bibinfo {author}
  {G.-Y. Chen},\ and\ \bibinfo {author} {F.~Nori},\ \emph {\bibinfo {title}
  {Einstein's Quantum Elevator: Hermitization of Non-Hermitian Hamiltonians via
  the Vielbein Formalism}},\ \href {\doibase 10.1103/physrevresearch.4.023070}
  {\bibfield  {journal} {\bibinfo  {journal} {Phys. Rev. Research}\ }\textbf
  {\bibinfo {volume} {4}},\ \bibinfo {pages} {023070} (\bibinfo {year}
  {2022})}\BibitemShut {NoStop}%
\bibitem [{\citenamefont {Ju}\ \emph {et~al.}(2019)\citenamefont {Ju},
  \citenamefont {Miranowicz}, \citenamefont {Chen},\ and\ \citenamefont
  {Nori}}]{Ju2019}%
  \BibitemOpen
  \bibinfo {author} {C.-Y. Ju}, \bibinfo {author} {A.~Miranowicz}, \bibinfo
  {author} {G.-Y. Chen},\ and\ \bibinfo {author} {F.~Nori},\ \emph {\bibinfo
  {title} {Non-Hermitian Hamiltonians and no-go theorems in quantum
  information}},\ \href {\doibase 10.1103/physreva.100.062118} {\bibfield
  {journal} {\bibinfo  {journal} {Phys. Rev. A}\ }\textbf {\bibinfo {volume}
  {100}},\ \bibinfo {pages} {062118} (\bibinfo {year} {2019})}\BibitemShut
  {NoStop}%
\bibitem [{\citenamefont {Misner}\ \emph {et~al.}(2017)\citenamefont {Misner},
  \citenamefont {Thorne},\ and\ \citenamefont {Wheeler}}]{Misner2017}%
  \BibitemOpen
  \bibinfo {author} {C.~W. Misner}, \bibinfo {author} {K.~S. Thorne},\ and\
  \bibinfo {author} {J.~A. Wheeler},\ \href {\doibase 10.2307/j.ctv301gk5}
  {\emph {\bibinfo {title} {Gravitation}}}\ (\bibinfo  {publisher} {Princeton
  University Press},\ \bibinfo {year} {2017})\BibitemShut {NoStop}%
\bibitem [{\citenamefont {Wald}(1984)}]{Wald}%
  \BibitemOpen
  \bibinfo {author} {R.~M. Wald},\ \href {\doibase
  10.7208/chicago/9780226870373.001.0001} {\emph {\bibinfo {title} {General
  Relativity}}}\ (\bibinfo  {publisher} {The University of Chicago Press},\
  \bibinfo {year} {1984})\BibitemShut {NoStop}%
\bibitem [{\citenamefont {Stoker}\ and\ \citenamefont
  {Carroll}(2019)}]{DonaldStoker2019}%
  \BibitemOpen
  \bibinfo {author} {D.~Stoker}\ and\ \bibinfo {author} {S.~M. Carroll},\ \href
  {\doibase 10.1017/9781108770385} {\emph {\bibinfo {title} {Spacetime and
  Geometry}}}\ (\bibinfo  {publisher} {Cambridge University Press},\ \bibinfo
  {year} {2019})\BibitemShut {NoStop}%
\bibitem [{\citenamefont {Collier}(2021)}]{Collier2021}%
  \BibitemOpen
  \bibinfo {author} {P.~Collier},\ \href {\doibase 10.4324/9781003444145-22}
  {\emph {\bibinfo {title} {A Beginner's Guide to Differential Forms}}}\
  (\bibinfo  {publisher} {Incomprehensible Books},\ \bibinfo {year} {2021})\
  pp.\ \bibinfo {pages} {311--311}\BibitemShut {NoStop}%
\bibitem [{\citenamefont {Needham}(2021)}]{Needham2021}%
  \BibitemOpen
  \bibinfo {author} {T.~Needham},\ \href {\doibase 10.1515/9780691219899}
  {\emph {\bibinfo {title} {Visual Differential Geometry and Forms}}}\
  (\bibinfo  {publisher} {Princeton University Press},\ \bibinfo {year}
  {2021})\BibitemShut {NoStop}%
\bibitem [{\citenamefont {Emam}(2021)}]{Emam2021}%
  \BibitemOpen
  \bibinfo {author} {M.~H. Emam},\ \href {\doibase
  10.1093/oso/9780198864899.001.0001} {\emph {\bibinfo {title} {Covariant
  Physics}}}\ (\bibinfo  {publisher} {Oxford University Press},\ \bibinfo
  {year} {2021})\BibitemShut {NoStop}%
\bibitem [{\citenamefont {Sakurai}\ and\ \citenamefont
  {Napolitano}(2017)}]{Sakurai2017}%
  \BibitemOpen
  \bibinfo {author} {J.~J. Sakurai}\ and\ \bibinfo {author} {J.~Napolitano},\
  \href {\doibase 10.1017/9781108499996} {\emph {\bibinfo {title} {Modern
  Quantum Mechanics}}}\ (\bibinfo  {publisher} {Cambridge University Press},\
  \bibinfo {year} {2017})\BibitemShut {NoStop}%
\bibitem [{\citenamefont {Mehri-Dehnavi}\ and\ \citenamefont
  {Mostafazadeh}(2008)}]{Mehri2008}%
  \BibitemOpen
  \bibinfo {author} {H.~Mehri-Dehnavi}\ and\ \bibinfo {author}
  {A.~Mostafazadeh},\ \emph {\bibinfo {title} {Geometric phase for
  non-Hermitian Hamiltonians and its holonomy interpretation}},\ \href
  {\doibase 10.1063/1.2968344} {\bibfield  {journal} {\bibinfo  {journal} {J.
  Math. Phys.}\ }\textbf {\bibinfo {volume} {49}},\ \bibinfo {pages} {082105}
  (\bibinfo {year} {2008})}\BibitemShut {NoStop}%
\bibitem [{\citenamefont {Nakahara}(2003)}]{Nakahara2003}%
  \BibitemOpen
  \bibinfo {author} {M.~Nakahara},\ \href {\doibase 10.1201/9781315275826-7}
  {\emph {\bibinfo {title} {Geometry, Topology and Physics}}},\ \bibinfo
  {edition} {2nd}\ ed.\ (\bibinfo  {publisher} {{IOP Publishing, Bristol}},\
  \bibinfo {year} {2003})\ pp.\ \bibinfo {pages} {244--307}\BibitemShut
  {NoStop}%
\bibitem [{\citenamefont {Xiao}\ \emph {et~al.}(2010)\citenamefont {Xiao},
  \citenamefont {Chang},\ and\ \citenamefont {Niu}}]{Xiao2010}%
  \BibitemOpen
  \bibinfo {author} {D.~Xiao}, \bibinfo {author} {M.-C. Chang},\ and\ \bibinfo
  {author} {Q.~Niu},\ \emph {\bibinfo {title} {Berry phase effects on
  electronic properties}},\ \href {\doibase 10.1103/RevModPhys.82.1959}
  {\bibfield  {journal} {\bibinfo  {journal} {Rev. Mod. Phys.}\ }\textbf
  {\bibinfo {volume} {82}},\ \bibinfo {pages} {1959} (\bibinfo {year}
  {2010})}\BibitemShut {NoStop}%
\bibitem [{\citenamefont {Wang}\ \emph {et~al.}(2015)\citenamefont {Wang},
  \citenamefont {Liu}, \citenamefont {Imri{\v{s}}ka}, \citenamefont {Ma},\ and\
  \citenamefont {Troyer}}]{Wang2015}%
  \BibitemOpen
  \bibinfo {author} {L.~Wang}, \bibinfo {author} {Y.-H. Liu}, \bibinfo {author}
  {J.~Imri{\v{s}}ka}, \bibinfo {author} {P.~N. Ma},\ and\ \bibinfo {author}
  {M.~Troyer},\ \emph {\bibinfo {title} {Fidelity Susceptibility Made Simple: A
  Unified Quantum Monte~Carlo Approach}},\ \href {\doibase
  10.1103/physrevx.5.031007} {\bibfield  {journal} {\bibinfo  {journal} {Phys.
  Rev. X}\ }\textbf {\bibinfo {volume} {5}},\ \bibinfo {pages} {031007}
  (\bibinfo {year} {2015})}\BibitemShut {NoStop}%
\bibitem [{\citenamefont {Tzeng}\ \emph {et~al.}(2021)\citenamefont {Tzeng},
  \citenamefont {Ju}, \citenamefont {Chen},\ and\ \citenamefont
  {Huang}}]{Tzeng2021}%
  \BibitemOpen
  \bibinfo {author} {Y.-C. Tzeng}, \bibinfo {author} {C.-Y. Ju}, \bibinfo
  {author} {G.-Y. Chen},\ and\ \bibinfo {author} {W.-M. Huang},\ \emph
  {\bibinfo {title} {Hunting for the non-Hermitian exceptional points with
  fidelity susceptibility}},\ \href {\doibase 10.1103/PhysRevResearch.3.013015}
  {\bibfield  {journal} {\bibinfo  {journal} {Phys. Rev. Res.}\ }\textbf
  {\bibinfo {volume} {3}},\ \bibinfo {pages} {013015} (\bibinfo {year}
  {2021})}\BibitemShut {NoStop}%
\bibitem [{\citenamefont {Tu}\ \emph {et~al.}(2022)\citenamefont {Tu},
  \citenamefont {Jang}, \citenamefont {Chang},\ and\ \citenamefont
  {Tzeng}}]{Tu2022}%
  \BibitemOpen
  \bibinfo {author} {Y.-T. Tu}, \bibinfo {author} {I.~Jang}, \bibinfo {author}
  {P.-Y. Chang},\ and\ \bibinfo {author} {Y.-C. Tzeng},\ \emph {\bibinfo
  {title} {General properties of fidelity in non-Hermitian quantum systems with
  {$\cal{PT}$} symmetry}},\ \href {\doibase 10.22331/q-2023-03-23-960}
  {\bibfield  {journal} {\bibinfo  {journal} {Quantum}\ }\textbf {\bibinfo
  {volume} {7}},\ \bibinfo {pages} {960} (\bibinfo {year} {2022})}\BibitemShut
  {NoStop}%
\bibitem [{\citenamefont {Nash}\ and\ \citenamefont {Sen}(2011)}]{Nash2011}%
  \BibitemOpen
  \bibinfo {author} {C.~Nash}\ and\ \bibinfo {author} {S.~Sen},\ \href
  {\doibase 10.1142/9599} {\emph {\bibinfo {title} {Topology and Geometry for
  Physicists}}}\ (\bibinfo  {publisher} {Dover Pub., New York},\ \bibinfo
  {year} {2011})\BibitemShut {NoStop}%
\bibitem [{\citenamefont {Polchinski}(1998)}]{Polchinski1998}%
  \BibitemOpen
  \bibinfo {author} {J.~Polchinski},\ \href {\doibase 10.1017/cbo9780511816079}
  {\emph {\bibinfo {title} {String Theory}}}\ (\bibinfo  {publisher} {Cambridge
  University Press},\ \bibinfo {year} {1998})\BibitemShut {NoStop}%
\bibitem [{\citenamefont {Becker}\ \emph {et~al.}(2006)\citenamefont {Becker},
  \citenamefont {Becker},\ and\ \citenamefont {Schwarz}}]{Becker2006}%
  \BibitemOpen
  \bibinfo {author} {K.~Becker}, \bibinfo {author} {M.~Becker},\ and\ \bibinfo
  {author} {J.~H. Schwarz},\ \href {\doibase 10.1017/cbo9780511816086} {\emph
  {\bibinfo {title} {String Theory and M-Theory}}}\ (\bibinfo  {publisher}
  {Cambridge University Press},\ \bibinfo {year} {2006})\BibitemShut {NoStop}%
\bibitem [{\citenamefont {Stefano}\ \emph {et~al.}(2019)\citenamefont
  {Stefano}, \citenamefont {Settineri}, \citenamefont {Macr{\`{\i}}},
  \citenamefont {Garziano}, \citenamefont {Stassi}, \citenamefont {Savasta},\
  and\ \citenamefont {Nori}}]{Stefano2019}%
  \BibitemOpen
  \bibinfo {author} {O.~D. Stefano}, \bibinfo {author} {A.~Settineri}, \bibinfo
  {author} {V.~Macr{\`{\i}}}, \bibinfo {author} {L.~Garziano}, \bibinfo
  {author} {R.~Stassi}, \bibinfo {author} {S.~Savasta},\ and\ \bibinfo {author}
  {F.~Nori},\ \emph {\bibinfo {title} {Resolution of gauge ambiguities in
  ultrastrong-coupling cavity quantum electrodynamics}},\ \href {\doibase
  10.1038/s41567-019-0534-4} {\bibfield  {journal} {\bibinfo  {journal} {Nat.
  Phys.}\ }\textbf {\bibinfo {volume} {15}},\ \bibinfo {pages} {803} (\bibinfo
  {year} {2019})}\BibitemShut {NoStop}%
\bibitem [{\citenamefont {Garziano}\ \emph {et~al.}(2020)\citenamefont
  {Garziano}, \citenamefont {Settineri}, \citenamefont {Stefano}, \citenamefont
  {Savasta},\ and\ \citenamefont {Nori}}]{Garziano2020}%
  \BibitemOpen
  \bibinfo {author} {L.~Garziano}, \bibinfo {author} {A.~Settineri}, \bibinfo
  {author} {O.~D. Stefano}, \bibinfo {author} {S.~Savasta},\ and\ \bibinfo
  {author} {F.~Nori},\ \emph {\bibinfo {title} {Gauge invariance of the Dicke
  and Hopfield models}},\ \href {\doibase 10.1103/physreva.102.023718}
  {\bibfield  {journal} {\bibinfo  {journal} {Phys. Rev. A}\ }\textbf {\bibinfo
  {volume} {102}},\ \bibinfo {pages} {023718} (\bibinfo {year}
  {2020})}\BibitemShut {NoStop}%
\bibitem [{\citenamefont {Settineri}\ \emph {et~al.}(2021)\citenamefont
  {Settineri}, \citenamefont {Stefano}, \citenamefont {Zueco}, \citenamefont
  {Hughes}, \citenamefont {Savasta},\ and\ \citenamefont
  {Nori}}]{Settineri2021}%
  \BibitemOpen
  \bibinfo {author} {A.~Settineri}, \bibinfo {author} {O.~D. Stefano}, \bibinfo
  {author} {D.~Zueco}, \bibinfo {author} {S.~Hughes}, \bibinfo {author}
  {S.~Savasta},\ and\ \bibinfo {author} {F.~Nori},\ \emph {\bibinfo {title}
  {Gauge freedom, quantum measurements, and time-dependent interactions in
  cavity {QED}}},\ \href {\doibase 10.1103/physrevresearch.3.023079} {\bibfield
   {journal} {\bibinfo  {journal} {Phys. Rev. Research}\ }\textbf {\bibinfo
  {volume} {3}},\ \bibinfo {pages} {023079} (\bibinfo {year}
  {2021})}\BibitemShut {NoStop}%
\bibitem [{\citenamefont {Savasta}\ \emph {et~al.}(2021)\citenamefont
  {Savasta}, \citenamefont {Stefano}, \citenamefont {Settineri}, \citenamefont
  {Zueco}, \citenamefont {Hughes},\ and\ \citenamefont {Nori}}]{Savasta2021}%
  \BibitemOpen
  \bibinfo {author} {S.~Savasta}, \bibinfo {author} {O.~D. Stefano}, \bibinfo
  {author} {A.~Settineri}, \bibinfo {author} {D.~Zueco}, \bibinfo {author}
  {S.~Hughes},\ and\ \bibinfo {author} {F.~Nori},\ \emph {\bibinfo {title}
  {Gauge principle and gauge invariance in two-level systems}},\ \href
  {\doibase 10.1103/physreva.103.053703} {\bibfield  {journal} {\bibinfo
  {journal} {Phys. Rev. A}\ }\textbf {\bibinfo {volume} {103}},\ \bibinfo
  {pages} {053703} (\bibinfo {year} {2021})}\BibitemShut {NoStop}%
\bibitem [{\citenamefont {Salmon}\ \emph {et~al.}(2022)\citenamefont {Salmon},
  \citenamefont {Gustin}, \citenamefont {Settineri}, \citenamefont {Stefano},
  \citenamefont {Zueco}, \citenamefont {Savasta}, \citenamefont {Nori},\ and\
  \citenamefont {Hughes}}]{Salmon2022}%
  \BibitemOpen
  \bibinfo {author} {W.~Salmon}, \bibinfo {author} {C.~Gustin}, \bibinfo
  {author} {A.~Settineri}, \bibinfo {author} {O.~D. Stefano}, \bibinfo {author}
  {D.~Zueco}, \bibinfo {author} {S.~Savasta}, \bibinfo {author} {F.~Nori},\
  and\ \bibinfo {author} {S.~Hughes},\ \emph {\bibinfo {title}
  {Gauge-independent emission spectra and quantum correlations in the
  ultrastrong coupling regime of open system cavity-{QED}}},\ \href {\doibase
  10.1515/nanoph-2021-0718} {\bibfield  {journal} {\bibinfo  {journal} {P. Soc.
  Photo-opt. Ins.}\ }\textbf {\bibinfo {volume} {11}},\ \bibinfo {pages} {1573}
  (\bibinfo {year} {2022})}\BibitemShut {NoStop}%
\bibitem [{\citenamefont {Born}\ and\ \citenamefont {Fock}(1928)}]{Born1928}%
  \BibitemOpen
  \bibinfo {author} {M.~Born}\ and\ \bibinfo {author} {V.~Fock},\ \emph
  {\bibinfo {title} {Beweis des Adiabatensatzes}},\ \href {\doibase
  10.1007/bf01343193} {\bibfield  {journal} {\bibinfo  {journal} {Z. Phys.}\
  }\textbf {\bibinfo {volume} {51}},\ \bibinfo {pages} {165} (\bibinfo {year}
  {1928})}\BibitemShut {NoStop}%
\bibitem [{\citenamefont {Berry}(1984)}]{Berry1984}%
  \BibitemOpen
  \bibinfo {author} {M.~V. Berry},\ \emph {\bibinfo {title} {Quantal Phase
  Factors Accompanying Adiabatic Changes}},\ \href {\doibase
  10.1142/9789813221215_0006} {\bibfield  {journal} {\bibinfo  {journal} {Proc.
  Royal Soc. London A}\ }\textbf {\bibinfo {volume} {392}},\ \bibinfo {pages}
  {45} (\bibinfo {year} {1984})}\BibitemShut {NoStop}%
\bibitem [{\citenamefont {Nandy}\ \emph {et~al.}(2018)\citenamefont {Nandy},
  \citenamefont {Taraphder},\ and\ \citenamefont {Tewari}}]{Nandy2018}%
  \BibitemOpen
  \bibinfo {author} {S.~Nandy}, \bibinfo {author} {A.~Taraphder},\ and\
  \bibinfo {author} {S.~Tewari},\ \emph {\bibinfo {title} {Berry phase theory
  of planar Hall effect in topological insulators}},\ \href {\doibase
  10.1038/s41598-018-33258-5} {\bibfield  {journal} {\bibinfo  {journal} {Sci.
  Rep.}\ }\textbf {\bibinfo {volume} {8}},\ \bibinfo {pages} {14983} (\bibinfo
  {year} {2018})}\BibitemShut {NoStop}%
\bibitem [{\citenamefont {Gu}(2010)}]{Gu2010}%
  \BibitemOpen
  \bibinfo {author} {S.-J. Gu},\ \emph {\bibinfo {title} {Fidelity approach to
  quantum phase transitions}},\ \href {\doibase 10.1142/s0217979210056335}
  {\bibfield  {journal} {\bibinfo  {journal} {International J. Mod. Phys. B}\
  }\textbf {\bibinfo {volume} {24}},\ \bibinfo {pages} {4371} (\bibinfo {year}
  {2010})}\BibitemShut {NoStop}%
\bibitem [{\citenamefont {Kato}(1976)}]{Kato1976}%
  \BibitemOpen
  \bibinfo {author} {T.~Kato},\ \href {\doibase 10.1007/978-3-642-66282-9_9}
  {\emph {\bibinfo {title} {Perturbation theory for linear operators}}},\
  \bibinfo {edition} {2nd}\ ed.,\ Grundlehren der mathematischen
  Wissenschaften\ (\bibinfo  {publisher} {Springer},\ \bibinfo {address}
  {Berlin},\ \bibinfo {year} {1976})\ pp.\ \bibinfo {pages}
  {479--515}\BibitemShut {NoStop}%
\bibitem [{\citenamefont {Heiss}(2004)}]{Heiss2004}%
  \BibitemOpen
  \bibinfo {author} {W.~D. Heiss},\ \emph {\bibinfo {title} {Exceptional points
  of non-{H}ermitian operators}},\ \href {\doibase 10.1088/0305-4470/37/6/034}
  {\bibfield  {journal} {\bibinfo  {journal} {J. Phys A: Math. Gen.}\ }\textbf
  {\bibinfo {volume} {37}},\ \bibinfo {pages} {2455} (\bibinfo {year}
  {2004})}\BibitemShut {NoStop}%
\bibitem [{\citenamefont {\"{O}zdemir}\ \emph {et~al.}(2019)\citenamefont
  {\"{O}zdemir}, \citenamefont {Rotter}, \citenamefont {Nori},\ and\
  \citenamefont {Yang}}]{Ozdemir2019}%
  \BibitemOpen
  \bibinfo {author} {{\c{S}}.~K. \"{O}zdemir}, \bibinfo {author} {S.~Rotter},
  \bibinfo {author} {F.~Nori},\ and\ \bibinfo {author} {L.~Yang},\ \emph
  {\bibinfo {title} {Parity{\textendash}time symmetry and exceptional points in
  photonics}},\ \href {\doibase 10.1038/s41563-019-0304-9} {\bibfield
  {journal} {\bibinfo  {journal} {Nat. Mater.}\ }\textbf {\bibinfo {volume}
  {18}},\ \bibinfo {pages} {783} (\bibinfo {year} {2019})}\BibitemShut
  {NoStop}%
\bibitem [{\citenamefont {Rattacaso}\ \emph {et~al.}(2020)\citenamefont
  {Rattacaso}, \citenamefont {Vitale},\ and\ \citenamefont
  {Hamma}}]{Rattacaso2020}%
  \BibitemOpen
  \bibinfo {author} {D.~Rattacaso}, \bibinfo {author} {P.~Vitale},\ and\
  \bibinfo {author} {A.~Hamma},\ \emph {\bibinfo {title} {Quantum geometric
  tensor away from equilibrium}},\ \href {\doibase 10.1088/2399-6528/ab9505}
  {\bibfield  {journal} {\bibinfo  {journal} {J. Phys. Commun.}\ }\textbf
  {\bibinfo {volume} {4}},\ \bibinfo {pages} {055017} (\bibinfo {year}
  {2020})}\BibitemShut {NoStop}%
\bibitem [{\citenamefont {Freedman}\ \emph {et~al.}(1976)\citenamefont
  {Freedman}, \citenamefont {van Nieuwenhuizen},\ and\ \citenamefont
  {Ferrara}}]{Freedman1976}%
  \BibitemOpen
  \bibinfo {author} {D.~Z. Freedman}, \bibinfo {author} {P.~van
  Nieuwenhuizen},\ and\ \bibinfo {author} {S.~Ferrara},\ \emph {\bibinfo
  {title} {Progress toward a theory of supergravity}},\ \href {\doibase
  10.1103/physrevd.13.3214} {\bibfield  {journal} {\bibinfo  {journal} {Phys.
  Rev. D}\ }\textbf {\bibinfo {volume} {13}},\ \bibinfo {pages} {3214}
  (\bibinfo {year} {1976})}\BibitemShut {NoStop}%
\bibitem [{\citenamefont {van Nieuwenhuizen}(1981)}]{Nieuwenhuizen1981}%
  \BibitemOpen
  \bibinfo {author} {P.~van Nieuwenhuizen},\ \emph {\bibinfo {title}
  {Supergravity}},\ \href {\doibase 10.1016/0370-1573(81)90157-5} {\bibfield
  {journal} {\bibinfo  {journal} {Phys. Rep.}\ }\textbf {\bibinfo {volume}
  {68}},\ \bibinfo {pages} {189} (\bibinfo {year} {1981})}\BibitemShut
  {NoStop}%
\bibitem [{\citenamefont {Kofman}\ \emph {et~al.}(2023)\citenamefont {Kofman},
  \citenamefont {Ivakhnenko}, \citenamefont {Shevchenko},\ and\ \citenamefont
  {Nori}}]{Kofman2023}%
  \BibitemOpen
  \bibinfo {author} {P.~O. Kofman}, \bibinfo {author} {O.~V. Ivakhnenko},
  \bibinfo {author} {S.~N. Shevchenko},\ and\ \bibinfo {author} {F.~Nori},\
  \emph {\bibinfo {title} {Majorana’s approach to nonadiabatic transitions
  validates the adiabatic-impulse approximation}},\ \href {\doibase
  10.1038/s41598-023-31084-y} {\bibfield  {journal} {\bibinfo  {journal} {Sci.
  Rep.}\ }\textbf {\bibinfo {volume} {13}},\ \bibinfo {pages} {5053} (\bibinfo
  {year} {2023})}\BibitemShut {NoStop}%
\end{thebibliography}%
\end{document}